\newif\ifsubmode
\submodefalse

\newif\ifprintfig
\printfigtrue

\ifsubmode
  \documentstyle[12pt,aasms4,epsf]{article}
  \received{}
  \accepted{}
  \journalid{}{}
  \articleid{}{}
\else
  \documentstyle[11pt,aaspp4,epsf]{article}
  \slugcomment{{\it Astronomical Journal, December 2002, in press}}
\fi

\lefthead{Gerssen et al.}
\righthead{Central Kinematics of M15 --- II.~Models}

\newcommand{\etal}{{et al.~}}

\newcommand{\lta}{\lesssim}
\newcommand{\gta}{\gtrsim}

\newcommand{\kms}{\>{\rm km}\,{\rm s}^{-1}}
\newcommand{\pc}{\>{\rm pc}}
\newcommand{\Msun}{\>{\rm M_{\odot}}}
\newcommand{\Lsun}{\>{\rm L_{\odot}}}

\begin{document}

\title{Hubble Space Telescope Evidence for an Intermediate-Mass Black
Hole in the Globular Cluster M15---\\ II.~Kinematical Analysis and
Dynamical Modeling\altaffilmark{1}}

\author{Joris Gerssen, Roeland P.~van der Marel}
\affil{Space Telescope Science Institute, 3700 San Martin Drive,
       Baltimore, MD 21218}

\author{Karl Gebhardt}
\affil{Astronomy Department, University of Texas at Austin, Mail Code C1400,
       Austin, TX 78712}

\author{Puragra Guhathakurta, Ruth C.~Peterson\altaffilmark{2}}
\affil{UCO/Lick Observatory, Department of Astronomy and Astrophysics,
       University of California at Santa Cruz, 1156 High Street,
       Santa Cruz, CA 95064}

\author{Carlton Pryor}
\affil{Department of Physics and Astronomy, Rutgers University, 136
       Frelinghuysen Road, Piscataway, NJ 08854-8019}

%%%%%%%%%%%%%%
% Additional affiliations
%%%%%%%%%%%%%%

\altaffiltext{1}{Based on observations made with the NASA/ESA Hubble Space
Telescope, obtained at the Space Telescope Science Institute, which is
operated by the Association of Universities for Research in Astronomy,
Inc., under NASA contract NAS 5-26555. These observations are
associated with proposal \#8262.}

\altaffiltext{2}{Also at: Astrophysical Advances, Palo Alto, CA 94301.}

%%%%%%%%%%%%%%%
% Start the abstract on a fresh page
%%%%%%%%%%%%%%%

\ifsubmode\else
\clearpage\fi

%%%%%%%%%%%%%%%
% Use a small baselineskip, unless in submission mode.
%%%%%%%%%%%%%%%

\ifsubmode\else
\baselineskip=14pt
\fi

%%%%%%%%%%%%%%%
% Abstract
%%%%%%%%%%%%%%%

\begin{abstract}
We have used the Space Telescope Imaging Spectrograph (STIS) on the
Hubble Space Telescope (HST) to obtain high spatial resolution
spectroscopy of the central region of the dense globular cluster M15.
The observational strategy and data reduction were described in
Paper~I (van der Marel \etal 2002). Here we analyze the extracted
spectra with a cross-correlation technique to determine the
line-of-sight velocities of individual stars. Our final STIS velocity
sample contains 64 stars, two-thirds of which have their velocity
measured for the first time. The new data set triples the number of
stars with measured velocities in the central projected $R \leq 1''$
of M15 and doubles the number in the central $R \leq 2''$. We combine
our data with existing ground-based data to obtain non-parametric
estimates of the radial profiles of the projected rotation velocity,
velocity dispersion, and RMS velocity $\sigma_{\rm RMS}$. The results
differ from earlier work in the central few arcsec in that we find
that $\sigma_{\rm RMS}$ rises to $\sim 14 \kms$, somewhat higher than
the values of $10$--$12 \kms$ inferred previously from ground-based
data.

To interpret the results we construct dynamical models based on the
Jeans equation for a spherical system. If the velocity distribution is
isotropic, then M15 must have a central concentration of non-luminous
material. If this is due to a single black hole, then a fit to the
full velocity information as function of radius implies that its mass
is $M_{\rm BH} = (3.9 \pm 2.2) \times 10^3 \Msun$. The existence of
intermediate-mass black holes in globular clusters is consistent with
several scenarios for globular cluster evolution proposed in the
literature. The inferred mass for M15 is consistent with the
extrapolation of the relation between $M_{\rm BH}$ and $\sigma_{\rm
RMS}$ that has been established for galaxies. Therefore, these results
may have important implications for our understanding of the evolution
of globular clusters, the growth of black holes, the connection
between globular cluster and galaxy formation, and the nature of the
recently discovered `ultra-luminous' X-ray sources in nearby galaxies.
Instead of a single intermediate-mass black hole, M15 could have a
central concentration of dark remnants (e.g., neutron stars) due to
mass segregation. However, we argue that the best-fitting
Fokker-Planck models that have previously been constructed for M15 do
not predict a central mass concentration that is sufficient to explain
the observed kinematics. To fit the M15 data without any central dark
mass concentration one must assume that the velocity distribution is
significantly radially anisotropic near the center, which contradicts
predictions from both Fokker-Planck and $N$-body calculations.
\end{abstract}

%%%%%%%%%%%%%%%
% Keywords
%%%%%%%%%%%%%%%

\keywords{globular clusters: individual (M15) ---
          stars: kinematics.}

\clearpage

%%%%%%%%%%%%%%%
% Beginning of main text
%%%%%%%%%%%%%%%

\section{Introduction}
\label{s:intro}

The globular cluster M15 (NGC 7078) has one of the highest central
densities of any globular cluster in our Galaxy. As a result, it has
been one of the globular clusters for which the structure and dynamics
have been most intensively studied in the past decade (as reviewed in
van der Marel 2001). The present paper is the second in a series of
two in which we present the results of a study with the Hubble Space
Telescope (HST) of the line-of-sight velocities of stars in the
central few arcsec of M15. Paper~I (van der Marel \etal 2002)
discussed the observations, and the extraction and calibration of the
stellar spectra. In the present paper we determine the stellar
line-of-sight velocities from the spectra, and we use the results to
study the dynamics and structure of M15.

M15 is a proto-typical core-collapsed cluster (Djorgovski \& King
1986; Lugger \etal 1987; Trager, King \& Djorgovski 1995), with a
stellar surface density profile that rises all the way into the
center. Such clusters make up $\sim\!  20$\% of all globular clusters
in our Galaxy, and stand in marked contrast to King-model clusters,
which show flat central cores and are modeled as tidally-truncated
isothermal systems. Even imaging studies with HST have not provided
any evidence for a homogeneous core in M15 (despite early claims to
the contrary; Lauer \etal 1991). Guhathakurta
\etal (1996) used the Second Wide Field and Planetary
Camera (WFPC2) and found the projected surface number density profile
inside $6''$ ($0.34$ pc) to be consistent with a power law $N(R)
\propto R^{-0.82 \pm 0.12}$. Sosin \& King (1997) used the Faint Object
Camera (FOC) and obtained $N(R) \propto R^{-0.70 \pm 0.05}$ for
turnoff stars. They also showed that the distributions for stars of
different masses have slightly different power-law slopes, which is
qualitatively consistent with the mass segregation predicted in a
cluster in which two-body relaxation has been important.

Bahcall \& Wolf (1976, 1977) constructed detailed models for the
equilibrium stellar density distribution of a globular cluster in
which a central black hole (BH) has been present for much longer than
the two-body relaxation time. For a cluster of equal-mass stars one
expects $N(R) \propto R^{-3/4}$, in surprisingly good agreement with
the observed star count profile for M15. While BHs have been
convincingly detected in the centers of galaxies (e.g., Kormendy \&
Gebhardt 2001), no convincing detections exist for globular
clusters. On the other hand, few, if any, previous studies have had
sufficient sensitivity to unambiguously detect BHs in globular
clusters with masses $M_{\rm BH} \lta 5 \times 10^3 \Msun$. There
are many ways in which globular cluster evolution at high densities
can lead to the formation of a massive BH in the center (Rees
1984). For example, core collapse induced by two-body relaxation may
lead to sufficiently high densities for individual stars or
stellar-mass black holes to interact or collide, with a single massive
BH as the likely end product (Sanders 1970; Quinlan \& Shapiro 1987,
1990; Lee 1987, 1993, 1995). Studies of such scenarios have gained
much interest lately (Miller \& Hamilton 2002; Mouri \& Taniguchi
2002; Portegies Zwart \& McMillan 2002) after the discovery of
intermediate luminosity X-ray objects in external galaxies (e.g.,
Colbert \& Mushotzky 1999). The emission of these objects may be due
to accretion onto intermediate mass BHs. However, this interpretation
is not uniquely implied by the data and there is no unique association
of these objects with star clusters (e.g., Zezas \& Fabbiano 2002).

While the observed star count profile of M15 is consistent with the
presence of a BH, it can be explained equally well as a result of
core-collapse (Grabhorn \etal 1992). Hence, the star count profile by
itself yields only limited insight. An additional problem is that
photometric studies cannot determine whether light follows mass, and
what the abundance and distribution of dark remnants are. Kinematical
studies are therefore essential to gain further insight. Integrated
light measurements of M15 initially suggested a very high central
dispersion, $\sigma = 25 \pm 7 \kms$. This was a sharp increase from
the dispersions of 5--$15 \kms$ found at larger radii from the radial
velocities of individual stars, which was interpreted as evidence for
the presence of a $10^3 \Msun$ central BH (Peterson, Seitzer \&
Cudworth 1989). This pioneering work spurred a lot of interest in
globular cluster dynamics. However, all subsequent studies were unable
to confirm the high central velocity dispersion measurement. In
particular, it became clear that that the weighting of stars by their
brightness in an integrated-light spectrum produces a large systematic
uncertainty in the velocity dispersion deduced from the broadening of
lines (Zaggia, Cappaccioli, \& Piotto 1993; Dubath \etal 1994).
Velocity measurements of individual stars are therefore called for.

Line-of-sight velocities are now known from ground-based studies for
$\sim\! 1800$ M15 stars, as compiled by Gebhardt
\etal (2000a). Many different studies contributed to
this dataset, as reviewed in the introduction of Paper~I.  The
projected velocity dispersion profile increases monotonically inwards
from $\sigma = 3 \pm 1 \kms$ at $R=7$ arcmin (Drukier \etal 1998), to
$\sigma = 11 \pm 1 \kms$ at $R=24''$. The analysis of Gebhardt \etal
(2000a) suggested that the velocity dispersion is approximately
constant at smaller radii, and is $\sigma = 11.7 \pm 2.8 \kms$ at the
innermost available radius $R \approx 1''$. Outside of the very
center, the velocity dispersion profile is well fitted by a spherical
dynamical model with an isotropic velocity distribution and a constant
mass-to-light ratio $\Upsilon = 1.7$ (in solar V-band units). However,
this model underpredicts the velocity dispersion in the central $2$
arcsec. The fit can be improved by addition of a central black hole,
which causes the velocity dispersion to rise in Keplerian fashion as
$\sigma \propto R^{-1/2}$ towards the center of the cluster. The best
fit was obtained with a mass $M_{\rm BH} \approx 2 \times 10^3 \Msun$
(Gebhardt \etal 2000a). However, the ground-based M15 velocity
dispersion data can be fitted equally well with a model in which the
mass-to-light ratio $\Upsilon(r)$ of the stellar population increases
inwards to a value of $\sim 3$ in the center. This would not {\it a
priori} be implausible, since mass segregation would tend to
concentrate heavy dark remnants to the center of the cluster. Models
with an anisotropic velocity distribution may even be able to fit the
data with constant mass-to-light ratio and without a central black
hole. Higher spatial resolution data and more detailed modeling are
necessary to decide amongst these scenarios; this is the focus of the
present series of papers.

It has been known for some time (Gebhardt \etal 1994) that M15 has a
net projected rotation amplitude of $V_{\rm rot} \approx 2 \kms$ at
radii comparable to the half-light radius (about 1~arcmin).  More
recent work (Gebhardt \etal 2000a; these results were also suggested
by the integrated-light measurements of Peterson 1993) has revealed
that the rotation amplitude is larger at small radii: $V_{\rm rot} =
10.4 \pm 2.7 \kms$ for $R \leq 3.4''$, implying that $V_{\rm
rot}/\sigma \approx 1$ in this region. This large amplitude is
surprising because two-body relaxation should rapidly transfer net
angular momentum outward from such small radii (see the discussion in
Gebhardt \etal 2000a).  Even more surprising is that the position
angle of the projected rotation axis at small radii is $\sim
100^{\circ}$ different from that near the half-light radius.  Although
the large increase in the rotation amplitude at small radii may have
something to do with the presence of a central BH (Gebhardt \etal
2000a), the increase and change in position angle are not predicted by
any current theory of globular cluster dynamical evolution.

Phinney (1993) used an alternative argument to constrain the mass
distribution of M15. There are two millisecond pulsars in M15 at a
distance $R = 1.1''$ from the cluster center that have a negative
period derivative ${\dot P}$.  This must be due to acceleration by the
mean gravitational field of the cluster, since the pulsars are
expected to be spinning down intrinsically (positive ${\dot P}$). The
observed ${\dot P}$ values place a strict lower limit on the mass
enclosed within a projected radius of $R = 1.1''$. Combined with the
observed light profile this implies that the mass-to-light ratio must
increase centrally inwards. A similar pulsar acceleration study was
recently performed by d'Amico \etal (2002) for the cluster NGC 6752,
which suggests a central increase in mass-to-light ratio in this
cluster as well. For M15, Phinney (1993) obtained $\Upsilon > 2.1$ for
the total mass-to-light ratio within $R \leq 1.1''$, with a
statistically most likely value of $\Upsilon \approx 3.0$. These
results are consistent with the analysis of stellar kinematics (see
also Dull \etal 1997). Unfortunately, the pulsar data, like the
ground-based observations of the kinematics, does not constrain the
distribution of mass tightly enough to discriminate between the
effects of mass segregation and a central BH.

Tighter constraints on the distribution of mass near the center of M15
need observations of the kinematics with better angular resolution
than previous studies and observations from space can supply these.
So we started a project to use HST to determine more stellar
velocities close to the center of M15 (HST program GO-8262, PI: van
der Marel). As described in Paper I, we used the Space Telescope
Imaging Spectrograph (STIS) to obtain observations with the
$0.1''$-wide slit at 18 adjacent positions near the cluster
center. All spectra cover the wavelength range from 5073--5359{\AA},
which includes the Mg b triplet at $\sim 5175${\AA}. The resolution is
$0.276${\AA} per pixel, which corresponds to $15.86 \kms$. Extensive
reductions and calibrations were performed to extract spectra with
signal-to-noise ratio $S/N > 5.5$ per pixel for a total of 131
stars. The velocity calibration of the spectra was the most crucial
and difficult aspect of the data reduction. Corrections were necessary
for: drifts in the wavelength scale during an orbit; changes in the
velocity of HST as it orbits the Earth; and wavelength shifts induced
by the offsets of stars from the center of the slit. The analyses in
Paper~I indicate that the uncertainty in velocity scale caused by
residual calibration errors in the final spectra is $\sim 2.5 \kms$.

Here we analyze the 131 stellar spectra from Paper~I, and we show that
for 64 of them the quality is sufficient to obtain an accurate
line-of-sight velocity measurement. We use the results to obtain new
constraints on the dynamics and structure of M15. The paper is
organized as follows. In Section~\ref{s:extraction} we describe the
cross-correlation algorithm that we have used for the extraction of
line-of-sight velocities, including the choice of spectral
templates. We discuss the reliability of the results based on an
analysis of our STIS observations of a calibration star. In
Section~\ref{s:kinresults} we describe the application of the
cross-correlation algorithm to the STIS spectra of M15.  We describe
how we have corrected the inferred velocities for the effects of
crowding and blending. The reliability of the inferred velocities is
verified by comparison to ground-based data, for those stars for which
the latter are available. In Section~\ref{s:profiles} we infer the
velocity dispersion and rotation velocity profiles of M15 from the
combined HST and ground-based line-of-sight velocity samples. In
Section~\ref{s:dynamics} we present dynamical models to interpret the
results, and we discuss the implications for the dynamical structure
and mass distribution of M15. Section~\ref{s:conc} discusses and
summarizes the main conclusions.

\section{Extraction of Line-of-Sight Velocities}
\label{s:extraction}

\subsection{Algorithm}
\label{ss:algorithm}

Extracting line-of-sight velocities from stellar absorption line data
is usually accomplished by cross-correlating the spectral data with a
template spectrum of known velocity.  One of the most widely used
implementations of the cross-correlation method has been developed by
Tonry and Davis (1979).  Several non cross-correlation based methods
have also been developed in the past two decades. While these methods
are better suited to extract the full line-of-sight velocity
distributions from absorption line spectra, they do so at the cost of
requiring high signal-to-noise ratio spectra.  For the STIS data
analyzed here, the only kinematical quantity of interest is the
line-of-sight velocity.  These data are thus best suited to a
cross-correlation based analysis.

To this end, all line-of-sight velocities were derived in IRAF using
the task XCSAO.  XCSAO is part of the RVSAO package described
extensively in Kurtz \& Mink (1998).  XCSAO is essentially a refined
and updated version of the cross-correlation algorithm developed by
Tonry \& Davis (1979). XCSAO is a well-tested and often used task with
many attractive features.  It provides a handle on both the confidence
and the error of a particular velocity measurement by means of the $r$
statistic, developed originally by Tonry \& Davis (1979), which we
refer to here as $r_{\rm cc}$ (in essence, this statistic measures the
amplitude of the cross-correlation peak, divided by the amplitude of
the `average' peak expected from noise and template
mismatch). Following Kurtz \& Mink (1998), we chose to calibrate this
statistic empirically (see below). Another useful feature of XCSAO is
its batch mode option which considerably simplifies the task of having
to analyze large numbers of spectra (our STIS data set consists of
19200 apertures).

All spectra were reduced and wavelength calibrated with an adaptation
of the STIS reduction pipeline as described in Paper~I.  Before the
actual cross-correlation takes place, XCSAO performs the following
tasks: continuum removal, apodization and Fourier filtering.  Due to
undersampling of the line spread function (see Paper~I) the STIS
spectra are strongly undulated, i.e. there are low frequency flux
variations with wavelength that are not related to the intrinsic
continuum distribution. To roughly preserve the correct line ratios,
the continuum was removed by division rather than by subtraction. The
ends of the spectra were cosine tapered to avoid aliasing. Finally, a
Fourier bandpass filter was applied to remove both the low frequencies
(residual continuum variations) and the high frequencies (noise). The
software then calculates the cross-correlation function and finds the
strongest peak in a $200 \kms$ range centered on the systemic velocity
of M15 ($-107.5 \kms$).  The position of this peak is fitted with a
parabolic function to derive the velocity and its formal random error.

\subsection{Spectral Templates}
\label{ss:templates}

The strength of a cross-correlation peak depends largely on how well
the template spectrum matches the observed spectrum.  Template spectra
were therefore obtained of a cluster star at 40 arcsec from the center
of M15, which was also used for target acquisition purposes (see
Paper~I).  This star is a very bright giant and relatively high
signal-to-noise spectra were therefore readily obtained. This ensures
that uncertainties in the inferred velocities result mainly from the
noise in the M15 data and are not associated with the template (save
template mismatch).

Templates observed with the same instrumental setup as the science
observations have the advantage that the instrumental signature will
be similar in both spectra.  Any instrumental effect can therefore be
easily calibrated out. However, the largest instrumental effect is
usually a broadening of the cross-correlation function due to
instrumental broadening of the absorption lines.  This does not
influence the position of the cross-correlation peak and, hence, does
not affect velocity measurements.

In addition to the observed templates, a number of artificial templates
were created based on Kurucz models. A grid of 30 model stellar
atmospheres with varying temperatures (4000K to 8000K), metallicities
($-2.1$ dex to $-2.5$ dex) and surface gravities ($\log g$ from 0.5 to 2.0)
covering a wavelength range from 5000 to 5500 \AA \ were
calculated.  The metallicity range was chosen to bracket the current
best estimate of M15's metallicity.

Artificial templates have the benefit of having infinite
signal-to-noise ratios and they can be constructed to closely match
the spectral features of the M15 data.  They lack the instrumental
signature, but as noted above, this would mainly affect measurements
of the velocity dispersion (i.e., the line widths) which is irrelevant
to the data presented here.

\subsection{Reliability of the Line-of-Sight Velocities}
\label{ss:reliability}

In order to successfully attain our goal of measuring the stellar
velocity dispersion within the central arcseconds of M15, several
intricate corrections to the derived line-of-sight velocities had to
be made. These corrections stem from the motion of HST (which has an
orbital velocity of $7.5 \kms$) and the fact that the stars generally
do not fall in the center of the slit (the full width of the slit
corresponds to $26.5 \kms$ in the dispersion direction). To calibrate
the necessary corrections, 14 short exposure spectra of the bright
field star HD 122563 (F8IV, magnitude $V=6.2$) were obtained at offset
positions parallel and perpendicular to the slit. Both the corrections
and the calibration are described fully in Paper~I. In the remainder
of this paper it should be implicitly understood that these
corrections were applied to all velocities. The star HD 122563 was
chosen because its low metallicity, ${\rm [Fe/H]} = -2.65 \pm 0.2$
(Sneden \& Parthasarathy 1983), which is close to the metallicity of
M15, ${\rm [Fe/H]} \approx -2.22$ (Harris 1996).

An additional use of the two-dimensional long-slit data obtained for
HD~122563 is to calibrate the confidence limits on the results
obtained with the XCSAO cross correlation software. In each of the 14
long-slit spectra we extracted 21 one-dimensional spectra at different
positions along the slit. This yielded 294 spectra, each for a $0.1''
\times 0.1''$ aperture placed somewhere within $1.0''$ from the
position of the star on the sky. Because of the wings of the PSF, each
of these spectra contains a meaningful amount of light from the star.
For each of these 294 spectra the line-of-sight velocities and the
errors were derived using the cross-correlation routine XCSAO. In
addition, the $r_{\rm cc}$ statistic was derived, as well as the
average signal-to-noise ratio $S/N$ per pixel (determined from the
error frames delivered by the HST/STIS pipeline). All apertures were
correlated against all Kurucz model templates and also against the
acquisition star template. The results were found not to change
significantly from template to template. This is most likely due to
the rather limited wavelength range of the observed spectra, 250
\AA. In the end we adopted the template which yielded the marginally
highest value of $r_{\rm cc}$.

The inferred velocities were used to empirically calibrate the
confidence limits in a manner analogous to Kurtz \& Mink (1998). The
left panel of Figure~\ref{f:calibconf} shows the logarithm of the
absolute difference between the measured line-of-sight velocity and
the literature value of the velocity of HD 122563 versus the $r_{\rm
cc}$ statistic. A clear break in the distribution of points occurs at
$r_{\rm cc} \approx 2.5$. For $r_{\rm cc} \gtrsim 2.5$ all velocity
differences are only a few $\kms$ and are thus considered reliable.
Not surprisingly, for small values of $r_{\rm cc}$ the inferred
velocities become unreliable.  As an additional test of the
reliability we have constructed a similar diagram showing the
logarithm of the absolute velocity difference versus the
signal-to-noise ratio (right panel of Figure~\ref{f:calibconf}).  This
plot shows the same qualitative behavior as the $r_{\rm cc}$ statistic
plot. It suggests that reliable velocities are derived for an average
S/N $\gtrsim 5$ per pixel.

An optimal selection of the $r_{\rm cc}$ and S/N cutoff limits can be
made by applying both criteria simultaneously.  Imposing a lower limit
on $r_{\rm cc}$ of 2.0 (vertical line in the left panel) would include
very deviant points.  However, all of the deviant points have S/N
values smaller than 5.5 (these points are shown as filled circles in
the left panel of Figure~\ref{f:calibconf}).  A similar situation is
encountered in the right panel, where all the deviant points that are
beyond the S/N cutoff, S/N $\gtrsim 5.5$, have unacceptable $r_{\rm
cc}$ values (again shown as filled circles). Thus, in both panels all
points that are to the right of their adopted cutoffs and are shown as
open circles are considered acceptable.  With the adopted values of
$r_{\rm cc}$ = 2.0 and S/N = 5.5, no deviant points remain while the
number of reliably measured velocities is maximized.  Extending the
$r_{\rm cc}$ and S/N cutoffs downward as far as possible is by no
means critical for the calibration star itself but in M15 the number
of reliably measured stars is at a premium. Kurtz \& Mink (1998)
generally use more conservative values of $r_{\rm cc}$. However, they
do not apply the $r_{\rm cc}$ statistic and the S/N criterion
simultaneously to their data.

The additional checks discussed below also indicate that the values
adopted here yield reliable results for our STIS setup.

\section{Kinematical Results}
\label{s:kinresults}

\subsection{Raw Stellar Velocities}
\label{ss:velocities}

As described in Paper~I, the STIS data set allowed the extraction of
(one-dimensional) spectra with $S/N > 5.5$ for a total of 131 stars in
M15. We analyzed each of the extracted one-dimensional spectra with
the cross-correlation algorithm described in
Section~\ref{s:extraction}. Based on the analysis of the calibration
star discussed in Section~\ref{s:extraction}, we expect that the
inferred velocities are reliable for all spectra that yield a
cross-correlation statistic $r_{\rm cc} > 2$. As a sanity check, we
plot in Figure~\ref{f:clusterconf} the quantity $(v - v_{\rm sys}) /
\sigma$ as a function of $r_{\rm cc}$. Here $v_{\rm sys} \approx
-107.5 \kms$ is the systemic heliocentric velocity of M15 and $\sigma
\approx 12 \kms$ is a rough estimate of the stellar velocity
dispersion near the center (Gebhardt \etal 2000a). This figure is
analogous to Figure~\ref{f:calibconf}. It shows that there are no
inferred velocities with $r_{\rm cc} > 2$ that are implausible in view
of our understanding of the kinematics of M15. So there seems no
reason to mistrust any of the inferred velocities with $r_{\rm cc} >
2$, which confirms the results from the calibration star analysis.

The STIS data yield a reliable velocity result (i.e., $S/N > 5.5$ and
$r_{\rm cc} > 2$) for 64 of the 131 extracted one-dimensional spectra.
According to the prescription of Paper~I, the 131 extracted spectra
were constructed by combining one or more apertures.  It makes
essentially no difference, however, whether the velocities are derived
directly from these co-added spectra, or whether they are derived by
averaging the velocities of individually analyzed apertures. The 64
selected stars are listed in Table~\ref{t:results}. We will refer to
this sample as our `STIS velocity sample'.  The identification number
in the first column corresponds to the entry number in the HST/WFPC2
stellar catalog described and presented in Paper~I. The inferred
heliocentric velocity $v_{\rm obs}$ is listed in column~6, and its
error $\Delta v_{\rm obs}$ in column~7. These are the `raw' velocities
inferred from the spectra, which are not yet corrected for
blending. Blending corrections are discussed in Section
\ref{ss:blending} below.  The velocity error $\Delta v_{\rm obs}$
listed for each star is the quadratic sum of the formal error of the
cross-correlation result and the spectral velocity calibration
uncertainty. For the latter we adopted a value of $2.5
\kms$, based on the results and discussion of Paper~I. The analysis of
the calibration star data shown in Figure~\ref{f:calibconf} confirms
that there there are no systematic uncertainties in the data that
exceed this value.

Figure~\ref{f:cmd} shows the color-magnitude diagram (CMD) of M15 for the
stars in the HST/WFPC2 catalog discussed in Paper~I. The stars in
the STIS velocity sample are highlighted. These stars generally have a
magnitude in $V$ brighter than 18.5, although there is one star that
is as faint as $V=19.05$.  Red giant and sub-giant stars dominate the
sample. This is true not only because their number density is higher
than for bluer horizontal branch stars, but also because blue stars
often lack sufficiently strong stellar absorption lines to yield a
strong cross-correlation peak (Mayor 1980). Two stars in the STIS
velocity sample (ID numbers 3393 and 5846; encircled crosses in
Figure~\ref{f:cmd}) occupy a location close to the instability strip and
are probably RR Lyrae variables. RR Lyrae stars are velocity variables
that can have velocity excursions of $\pm 50 \kms$ from the mean
(Smith 1995). We therefore exclude these two stars from our subsequent
dynamical analysis.

The distance $d$ of the stars in the STIS velocity sample from the
cluster center ranges from 0.2 to 27 arcsec. The adopted position of
the cluster center is given in Paper~I, and is based on the analysis
of Guhathakurta \etal (1996). The influence of the uncertainty in this
position ($0.2''$ in each coordinate) on the results of our study is
negligible, as discussed in Section~\ref{ss:isoconstant}. The solid
line in the left panel of Figure~\ref{f:disthist} shows a histogram of
the distribution of the stars in the STIS velocity sample with
distance. For comparison, the dashed line shows the distribution of
the stars for which a velocity determination (with an uncertainty
better than $10 \kms$) was known from ground-based data (Gebhardt
\etal 2000a).  The logarithm of their corresponding cumulative
distributions are shown in the right panel.  The heavy solid line in
this panel shows the distribution for the combined STIS and
ground-based velocity samples. It is clear that the STIS data
significantly increase the number of stars with known velocities close
to the cluster center. Figure~\ref{f:xyplot} shows the spatial
distribution in the central $4'' \times 4''$ of the stars with known
line-of-sight velocities.

\subsection{Consistency Checks}
\label{ss:infvelocities}

About one third of the stars in the STIS velocity sample also have a
ground-based measurement of the line-of-sight velocity (Gebhardt \etal
2000a). A comparison of the ground-based velocities to the HST/STIS
velocities is shown in Figure~\ref{f:velcomp}. The agreement is
generally excellent, which confirms the reliability of the HST
results.  The residual velocities ($v_{\rm HST} - v_{\rm ground}$) are
shown in Figure~\ref{f:residuals} as a function of $V$-band magnitude
and $B-V$ color. There are no trends with either, which once more
validates the velocities determined with STIS.  The weighted mean
residual velocity is $0.2 \pm 1.4 \kms$, indicating that our velocity
zero point is consistent with what has been inferred from ground-based
observations.  The $\chi^2$ value is $29.1$ for 23 points, which is
statistically acceptable.

Neither the HST velocities nor the ground-based velocities in
Figure~\ref{f:velcomp} contain an explicit correction for
blending\footnote{The ground-based velocities cannot easily be
corrected for blending. So when comparing the ground-based velocities
to the HST velocities it makes sense to use HST velocities that are
not corrected for blending either. In the subsequent dynamical
analysis we do correct the HST velocities for blending, as described
in Section~\ref{ss:blending} below.}. However, even the best
ground-based data (Gebhardt \etal 2000a) has a poorer PSF than the HST
data, so some of the small discrepancies seen in
Figure~\ref{f:velcomp} may be due to the effects of blending. This is
true in particular for the two labeled stars in
Figure~\ref{f:velcomp}, which appear to lie somewhat offset from the
one-to-one relation. The first star, ID~5222, is close ($\sim 0.1''$)
to another star and it is possible that the ground-based determination
of its velocity is more strongly affected by blending than the HST
measurement. The second star, ID~5831, is a member of the triple star
system AC~214 that is located within $0.5''$ of the M15 center. Its
three $V \approx 15.5$ members are separated by about $0.1''$. Our
STIS sample contains the velocities of two of AC~214's members
(ID~5831 and ID~5846). We were unable to extract a reliable velocity
for the bluest of the three stars (ID~5872, $B-V =
0.02$). Interestingly, for ID~5831 and ID~5846 we find a velocity
difference of $\sim 50 \kms$. This provides a tantalizing hint for the
existence of a central compact object in the midst of AC~214.
However, ID~5846 is a possible RR Lyrae variable and the measured
velocity difference is consistent with the maximum velocity excursion
of an RR Lyrae variable.  The ground-based measurement ($-116 \kms$,
Gebhardt \etal 2000a) of ID~5831 is consistent with being the weighted
mean of the two values inferred from the HST/STIS data. The
ground-based measurement therefore probably suffered from blending of
light with ID~5846.

In addition to the comparison with ground-based results, we have also
performed internal consistency checks on the STIS data set. As
described in Paper I, the observational strategy yielded several
independent spectra for most stars. These independent spectra
correspond either to different rows in the same exposure (i.e.,
adjacent positions along the slit for a given slit position) or to
different exposures (i.e., adjacent slit positions on the sky,
observed either within the same telescope visit to the target or in
different visits). The strategy adopted for our final results has been
to co-add spectra for a given star, and to analyze the grand-total
summed spectrum. However, for testing we have also performed a
cross-correlation analysis of all 19,200 individual spectra in our
dataset. We extracted those results that conform to the selection
criteria with respect to blending, $S/N$ and $r_{\rm cc}$ that were
discussed previously. For many stars in our STIS velocity sample this
yields two or more independent velocity measurements. Let $v_i$ and
$v_j$ be two independent velocity measurements for the same star, and
let $\Delta v_i$ and $\Delta v_j$ be the corresponding
uncertainties. One then expects each difference $\Delta v_{ij} \equiv
v_i - v_j$ to be a Gaussian random deviate drawn from a distribution
with zero mean and dispersion $\epsilon_{ij} \equiv [(\Delta v_i)^2 +
(\Delta v_j)^2)]^{1/2}$. Figure~\ref{f:repeat} shows a histogram of
all the $\Delta v_{ij}$. It includes those stars for which multiple
independent measurements could be extracted; each star can appear many
times in the histogram, depending on how many different measurements
are available. The histogram is well fit by a Gaussian with a
dispersion of $6 \kms$ (solid curve). This is in reasonable agreement
with the expected value of $5 \kms$, based of the average of the
individual uncertainties $\epsilon_{ij}$. This implies that there is
no reason to suspect the presence of unidentified systematic errors in
the analysis, consistent with the results from the comparison of STIS
and ground-based results.

\subsection{Statistical Corrections for Blending}
\label{ss:blending}

Even with the resolution of HST, the light profiles of stars in the
densely populated central region of M15 overlap partially with
neighboring stars. So when the spectrum of an individual star is
extracted from our HST/STIS data set, only a fraction $f$ of the total
light is actually due to the star of interest. The remaining fraction
$(1-f)$ is due to `contaminating' stars. The algorithm for spectral
extraction employed in Paper~I was designed to yield values $f \geq
0.75$. We verified that the main results of our analysis are
insensitive to the exact choice of this cut (see also
Section~\ref{s:profiles} below). The fraction $f$ is listed in
Table~\ref{t:results} for each spectrum that was analyzed. The
blending fraction $(1-f) \leq 0.25$ is small, but does need to be
corrected for.

To correct for blending we note that the observed velocity $v_{\rm
obs}$ (the velocity extracted from the one-dimensional spectrum using
cross-correlation) is, to a good approximation, equal to the
luminosity weighted mean velocity of all the light that contributes to
the spectrum. Therefore,
\begin{equation}
\label{blendv}
  v_{\rm obs} = f v_{\rm star} + (1-f) v_{\rm cont} ,
\end{equation}
where $v_{\rm star}$ the velocity of the star of interest, and $v_{\rm
cont}$ is the luminosity weighted mean velocity of the contaminating
stars.  The analysis of Paper~I tells us which stars are contaminating
the spectrum and what fraction of the light they contribute. However,
we do not generally know the velocities of the contaminating stars. On
the other hand, we do know something about the contaminating stars:
they belong to the cluster. Hence, the velocities of the contaminating
stars are random deviates drawn from the line-of-sight velocity
distribution of the cluster. The expectation value of the statistical
quantity $v_{\rm cont}$ is therefore equal to $v_{\rm sys}$. The
dispersion is equal to $A \sigma$, where $\sigma$ is the line-of-sight
velocity dispersion of the cluster at the position of interest. The
constant $A \leq 1$ depends on the number of stars that provide the
contaminating light. If there is only a single star, then $A=1$. Upon
taking the first and second moments of equation~(\ref{blendv}) we can
calculate the expectation value of $v_{\rm star}$, which we denote
$v_{*}$, and the dispersion of $v_{\rm star}$, which we denote $\Delta
v_{*}$. This yields
\begin{equation}
\label{corrections}
  v_{*} =
        v_{\rm sys} + {1 \over f} [v_{\rm obs} - v_{\rm sys}] , \qquad
  \Delta v_{*} =
        {1 \over f} [ (\Delta v_{\rm obs})^2 + (1-f)^2 A^2 \sigma^2 ]^{1/2} ,
\end{equation}
where $\Delta v_{\rm obs}$ is the formal error of the velocity
inferred from the observed spectrum through cross-correlation. In the
following we assume that $A=1$; this is conservative in the sense that
it yields the largest uncertainty $\Delta v_{*}$.

The quantities $v_{*}$ and $\Delta v_{*}$ are the final estimate of
the stellar velocity and its error. They result from correcting the
measurement $v_{\rm obs}$ and its error $\Delta v_{\rm obs}$ for
blending. The accuracy of these corrections was verified through
Monte-Carlo simulations. Equation~(\ref{corrections}) shows that the
correction is such that it increases the deviation of the stellar
velocity from the cluster mean by a factor $1/f$. This is because
blending will, on average, tend to draw measured velocities towards
the cluster mean. Equation~(\ref{corrections}) also shows that the
error $\Delta v_{*}$ of the corrected velocity is larger than the
error $\Delta v_{\rm obs}$ of the measured velocity. This is so
because of our poor knowledge of the velocities of the contaminating
stars. The values of $v_{*}$ and $\Delta v_{*}$ are listed in
Table~\ref{t:results} for all the stars in the STIS velocity
sample. They do not depend sensitively on the exact choices of $v_{\rm
sys}$ and $\sigma$, for which we adopted $v_{\rm sys} = -107.5 \kms$
and $\sigma = 12 \kms$ (Gebhardt \etal 2000a).

The median value of $f$ in our velocity sample is $0.85$. The
uncorrected velocities have a median $|v_{\rm obs} - v_{\rm sys}|$ of
$8.8 \kms$ and a median error $\Delta v_{\rm obs}$ of $3.8 \kms$.  The
corrected velocities have a $|v_{*} - v_{\rm sys}|$ of $11.0
\kms$ and a median error $\Delta v_{*}$ of $5.4 \kms$. Thus, the
blending corrections are always relatively small. This is because
$(1-f)$ is always less than 25 percent; stars in our data set with
larger amounts of blending simply did not make it into our final STIS
velocity sample.

\section{Rotation Velocity and Velocity Dispersion Profiles}
\label{s:profiles}

The kinematical analysis of the data and the subsequent dynamical
modeling benefit from having the largest possible sample. We have
therefore combined the 64 stars in the STIS velocity sample with the
1777 stars in the ground-based sample used previously by Gebhardt
\etal (2000a). There are 23 stars in common to both samples (see
Table~\ref{t:results} and Figures~\ref{f:velcomp}
and~\ref{f:residuals}). Those stars were assigned a velocity based on
the weighted mean of the two measurements\footnote{The 1-$\sigma$
error in a weighted average of $N$ measurements $z_i$ is given by
$[\sum_{i=1}^{N} 1 / (\Delta z_i)^2 ]^{-1/2}$. However, this
underestimates the true error if there are systematic errors in
addition to Gaussian random errors. These cases manifest themselves by
having a $\chi^2$ that is larger than the number of degrees of freedom
(where the $\chi^2$ is defined to measures the residuals of the data
with respect to their weighted average). In these cases we increased
the errors to yield a $\chi^2$ equal to the number of degrees of
freedom.}. One could argue that the STIS velocities should be more
accurate, so that it might be more appropriate to ignore the
ground-based velocity measurements for these stars. However, since the
HST and ground-based velocities are generally in good agreement
(Figure~\ref{f:residuals}), this doesn't affect any of our final
results. Based on their locations in the CMD diagram, a total of 19
stars (including two in the HST sample) are possible RR Lyrae
candidates. These stars were excluded from the sample, yielding a
final sample of 1797 stars.\footnote{We note that the ground-based
velocities in the sample were not corrected for blending in the same
way that the HST/STIS velocities were. This should not be a problem,
because the ground-based velocities dominate the sample only at radii
where crowding is less of a problem. We verified this with a number of
tests, for example, by studying the kinematics inferred from stars of
different magnitudes. From these tests we conclude that potential
blending in the ground-based data does not affect the main conclusions
of our paper at a significant level.}

We have used two non-parametric techniques to estimate kinematical
profiles from the discrete velocity measurements: {\tt LOWESS} and
{\tt SROTATE}. These techniques yield the velocity dispersion and the
amplitude and position angle of rotation, as functions of radius.
{\tt LOWESS} is a smoothing-based technique that uses a locally
weighted regression algorithm, and is described fully in Gebhardt
\etal (1994). It finds a smooth curve for the kinematical
profiles as a function of radius. At each radius, it uses a fixed
fraction of the data points in both radial directions. The data points
are weighted depending on their distance from the radius of interest
using an inverted parabolic function. The smoothing parameter is the
fraction of the data included in the fit at any given radius, for
which we generally use 10\%. The kinematical parameters are estimated
using a maximum likelihood estimator that includes both the
uncertainties on the individual velocity measurements and their
assigned weights. The statistical uncertainties on the inferred
profiles are calculated using Monte Carlo simulation. An undesirable
feature of {\tt LOWESS} is that it tends to wash out small scale
features, which can bias the kinematical profiles close to the
center. This problem is circumvented by the technique {\tt SROTATE},
which is described fully in Gebhardt \etal (2000a) (the technique is
not referred to as {\tt SROTATE} in that paper, but it is described in
the fifth paragraph of its section 3.1). It is similar to {\tt LOWESS}
in most respects, and differs primarily in that it allows for a
variable size of the data window (i.e., of the smoothing). At most
radii, it used a fixed fraction of the data points in both radial
directions, as does {\tt LOWESS}. However, at the boundaries of the
dataset the window size shrinks to a smaller number of data points,
although no smaller than a preset `minimum window size'. This approach
preserves sharp radial variations close to the cluster center, while
also maintaining the desirable feature of {\tt LOWESS} of lessening
large variations due to discrepant data points. The results described
in the remainder of this section were all obtained with {\tt SROTATE}
and a minimum window size of 11 data points. The main features of the
results were found not to depend sensitively on the exact choice of
the minimum window size. As a check we also determined the kinematical
profiles with {\tt LOWESS}. Although the results from {\tt LOWESS} are
somewhat unreliable at the smallest radii, it did yield results that
are statistically consistent with those derived from {\tt
SROTATE}. Overall, the kinematical profiles presented here are robust,
and do not depend sensitively on either the choice of the analysis
algorithm of the choice of the algorithm parameters.

Figure~\ref{f:kinprofile}a--c show the radial profiles of the velocity
dispersion $\sigma$, the rotation velocity $V_{\rm rot}$, and the
position angle ${\rm PA}_{\rm kin}$ of the kinematical major axis.
The $68.3$\% confidence bands are indicated with dotted curves. At
radii larger than a few arcsec the data set is dominated by stars with
ground-based velocity measurements and, as a consequence, the results
derived at these radii are almost identical to those obtained
previously by Gebhardt \etal (2000a, their Figure~13). Therefore, the
results at the smallest radii are of primary interest here. Gebhardt
\etal (2000a) found an increase in the rotation velocity of M15
towards the center of the cluster, with a maximum of $V_{\rm rot}
\approx 10 \kms$ at a radius $R \approx 2.5''$. We find that the
rotation velocity actually continues to rise inward to a value of
$V_{\rm rot} \approx 13 \kms$ at our innermost (trustworthy) radius,
$R \approx 0.5''$.  The velocity dispersion $\sigma$ shows a mild
inwards decrease at small radii, from $\sigma \approx 14 \kms$ at $R
\approx 4''$ to $\sigma \approx 10 \kms$ at $R \approx 0.5''$. This is
the same radial range over which the rotation velocity shows a strong
inwards increase. The value of ${\rm PA}_{\rm kin}$ appears to
decrease by $\sim 50^{\circ}$ when going inward from $R \approx 4''$
to $R \approx 0.5''$, but this decrease is only marginally
significant. The fact that $V_{\rm rot}/\sigma \approx 1$ near the
center of M15 is quite surprising.  This implications of this were
already discussed at length by Gebhardt \etal (2000a), and we will
therefore not comment further on it here.

The pressure in a stellar system is determined by the mean squared
velocity of the stars (Binney \& Tremaine 1987). In hydrostatic
equilibrium, the pressure gradient balances the gravitational force.
The quantity that is of primary importance from a dynamical point of
view is therefore neither $V_{\rm rot}$ nor $\sigma$, but the mean
squared projected line-of-sight velocity $V_{\rm rot}^2 +
\sigma^2$. We define $\sigma_{\rm RMS}(R) \equiv \langle V_{\rm
rot}^2 + \sigma^2 \rangle^{1/2}$, where the angle brackets denote the
average over a ring of radius $R$ on the projected plane of the
sky. The most convenient way to determine $\sigma_{\rm RMS}$ from the
discrete velocity measurements is to use {\tt SROTATE} while keeping
the rotation velocity fixed at zero. The resulting profile is shown in
Figure~\ref{f:kinprofile}d. Figure~\ref{f:trend} is similar, but for
visualization purposes it also shows the individually measured stellar
velocities for comparison. The quantity $\sigma_{\rm RMS}$ measures
the spread among these individual velocities (properly corrected for
the small contribution from measurement uncertainties). The
uncertainty in the determination of $\sigma_{\rm RMS}$ scales as the
inverse square-root of the number of data points, and is at all radii
much smaller than $\sigma_{\rm RMS}$ itself. Outside the central few
arcsec, the data are dominated by stars with ground-based velocity
measurements and the $\sigma_{\rm RMS}$ profile is almost identical to
that inferred by Gebhardt \etal (2000a). However, at smaller radii our
results are more accurate because of the addition of velocity
measurements from HST. We find that $\sigma_{\rm RMS}$ increases from
$\sim 10 \kms$ at $R = 10''$ to $\sim 14 \kms$ at $R = 3''$. Inside of
the central few arcsec $\sigma_{\rm RMS}$ is consistent with being
constant at $\sim 14 \kms$. This exceeds the central values of
$10$--$12 \kms$ inferred by Dull \etal (1997) and Gebhardt \etal
(2000a).

In our analysis we have included only those STIS spectra for which a
single star contributes a fraction $f$ of the total light that exceeds
$f_{\rm min} = 0.75$. If one chooses a more stringent limit, then the
number of stars for which velocities can be determined decreases. For
example, if one uses $f_{\rm min} = 0.90$, then the size of the STIS
velocity sample reduces to only 22 stars (see
Table~\ref{t:results}). On the other hand, these stars are the ones
for which the measurements are most secure. It is therefore a useful
test to check that the results for these stars are consistent with the
results for the sample as a whole. The STIS velocity sample contains
13 stars with $f \geq 0.90$ that reside at $R \leq 6''$ from the M15
center. These stars have an average distance from the cluster center
$\langle R \rangle = 3.3''$ and their RMS projected line-of-sight
velocity is $\sigma_{\rm RMS} = 14.1 \pm 3.2 \kms$. This is in
excellent agreement with the results obtained from the full sample,
shown in Figure~\ref{f:kinprofile}d. This indicates that main results
of our study do not depend significantly on either the amount of
blending allowed during the extraction of the STIS spectra (see
Paper~I) or the subsequent blending corrections (see
Section~\ref{ss:blending}).\footnote{The algorithm of Paper~I that
constructs a grand-total spectrum for each star has more parameters
than just $f_{\rm min}$ (although $f_{\rm min}$ is certainly the most
important one). For example, an individual aperture is rejected in the
construction of the grand-total spectrum for a star if that star does
not contribute at least a fraction $g > g_{\rm min} = 0.5$ of the
light in the aperture. As for $f_{\rm min}$, the results of our
analysis are not sensitive to the precise choice of $g_{\rm min}$. The
reason for this is that most of the light in the final grand-total
spectrum for each star always comes from apertures with $g
\gta f_{\rm min}$ (otherwise it would not be possible for the sum of the 
aperture spectra to have $f \geq f_{\rm min}$). Since we choose
$f_{\rm min} \gg g_{\rm min}$, the precise choice of $g_{\rm min}$ has
negligible influence on the final spectra.}

To assess the implications of our new observational results we proceed
in Section~\ref{s:dynamics} by constructing detailed dynamical
models. As discussed below, these models use a maximum likelihood
approach that operates directly on the individually observed stellar
velocities. This has the advantage that there is no dependence on
arbitrary binning or smoothing parameters, such as the data window
size in {\tt SROTATE}. Nonetheless, the kinematical profiles in
Figure~\ref{f:kinprofile} are important for providing insight into
both the dynamical state of M15 and the results of the likelihood
analysis (although the profiles are not themselves used in the
dynamical modeling).

\section{Dynamical Modeling}
\label{s:dynamics}

\subsection{Formalism}
\label{ss:modtechnique}

To determine the mass distribution in M15 we assume that the cluster
is spherical and that it is in hydrostatic equilibrium. The assumption
of sphericity is adequate because the isodensity contours of M15 are
not far from circular, at least at the small radii that are of primary
interest here (Guhathakurta \etal 1996). The assumption of hydrostatic
equilibrium is also adequate, despite the fact that a cluster like M15
undergoes secular evolution as a result of two-body relaxation. The
two-body relaxation time significantly exceeds the dynamical time and
the evolution therefore proceeds through a series of states that are
all approximately in hydrostatic equilibrium. A spherical system in
hydrostatic equilibrium obeys the Jeans equation (Binney \& Tremaine
1987):
\begin{equation}
\label{jeans}
  \frac{1}{\rho} \frac{d(\rho \sigma_r^2)}{dr} +
  \frac{2 \beta \sigma_r^2}{r} +
  \frac{d \Phi}{dr} = 0.
\end{equation}
Here $\rho$ is the mass density, $\Phi$ is the gravitational
potential, $\beta \equiv 1 - \sigma_t^2 / \sigma_r^2$ is a measure of
the anisotropy of the stellar velocity distribution, and $\sigma_r^2$
and $\sigma_t^2$ are the mean-squared velocities in the radial and
tangential directions, respectively. In general, all these quantities
are a function of the three-dimensional radius $r$. Note that the
assumption of sphericity does not imply that rotation is fully
ignored. It is possible to construct spherical dynamical models that
rotate. Rotation contributes to the mean-squared velocities in the
azimuthal direction, and therefore is included in
equation~(\ref{jeans}). So while we do not model the rotation of M15
explicitly, we do include its contribution to the hydrostatic support.

At a given position in three-dimensional space, the mean-squared
velocity in the direction along the line-of-sight is $\sigma_l^2 =
\sigma_r^2 [1 - \beta(r) R^2/r^2]$, where $R$ is the projected radius
on the sky. The calculation of the observable projected kinematical
quantities depends on whether one observes integrated light or
individual stars. In the integrated light situation, the mean-squared
projected line-of-sight velocity $\sigma_{\rm RMS}^2$ is the
line-of-sight projection of $j \sigma_l^2$, divided by the
line-of-sight projection of $j$. Here $j(r)$ is the three-dimensional
luminosity density, which upon line-of-sight projection yields $I(R)$,
the projected intensity. By contrast, for M15 we are in the discrete
situation, which is generally more complicated. Instead of $j(r)$ one
must know the three-dimensional number density of that population of
stars from which the stars with kinematical measurements are drawn.
This quantity is more difficult to estimate. To eliminate this
complication we assume throughout our analysis that the number density
of these stars is linearly proportional to the luminosity density
$j(r)$, with a proportionality constant that is independent of
position. So we assume that the average luminosity per observed star
does not depend on position. This is a minor simplification, because
it doesn't allow for gradients in the {\it visible} stellar population
within the cluster. Note, however, that no assumption is made about
the distribution of dark objects in the cluster, which is the much
more important quantity for the dynamical predictions of the modeling.

In practice, we start the modeling from the $V$-band surface
brightness profile $\mu_V(R)$ in mag/arcsec$^2$. For this we take the
profile compiled from various sources by Trager \etal (1995). We
correct this profile for $A_V = 0.28$ mag of extinction, which is
based on the reddening $E(B-V) = 0.09$ (Harris 1996) and a standard
Galactic reddening law with $A_V / E(B-V) = 3.1$ (Binney \& Merrifield
1998). We transform this profile into a projected intensity $I(R)$ in
$\Lsun \pc^{-2}$. The Trager \etal profile is based exclusively on
ground-based data. Because of this we do not use their profile inside
of $R = 5''$, where it is compromised by seeing and crowding. For $R
\leq 5''$ we use the projected number density profile derived
non-parametrically by Guhathakurta \etal (1996) from HST/WFPC2 data,
scaled to match the Trager et al profile at $R=5''$. This profile
rises into the center of M15 approximately as a power law, $I(R)
\propto R^{-0.82}$. Sosin \& King (1997) inferred a slightly different
profile from HST/FOC data, and they also found slightly different
power-law slopes for different stellar masses. Inside $\sim 0.3''$ the
number density profile is not well constrained by either study, due to
the limited number of stars and uncertainties in the position of the
M15 center. We have performed a variety of tests to assess the
influence of these uncertainties on the predictions of our dynamical
models. We have found that the main results are quite robust and do
not depend sensitively on the exact choice of $I(R)$ in the central $5''$.

To calculate isotropic ($\beta=0$) models for M15 we have used the
non-parametric technique of Gebhardt \& Fischer (1995). It solves an
Abel integral to calculate $j(r)$ from $I(R)$. The mass density is
then calculated as $\rho(r) = \Upsilon(r) j(r)$, where $\Upsilon(r)$
is an assumed mass-to-light ratio profile. Poisson's equation is then
solved to obtain the gravitational potential $\Phi(r)$. To this we add
the contribution $-G M_{\rm BH} / r$ of a possible central massive
black hole. The Jeans equation~(\ref{jeans}) is then solved, and the
results are projected along the line of sight to obtain $\sigma_{\rm
RMS}(R)$. To calculate anisotropic models ($\beta \not= 0$), we have
used the technique of van der Marel (1994). This technique starts with
a parametric fit to $I(R)$, but otherwise proceeds similarly as the
technique of Gebhardt \& Fischer (1995). As a test, we used both
techniques to calculate predictions for the case $\beta = 0$ and
$\Upsilon(r) = $ constant, which yielded excellent agreement. In all
model calculations we assumed a distance to M15 of 10.0 kpc (Harris
1996).

\subsection{Comparing the Data and Models}
\label{ss:compare}

Data-model comparison can be performed qualitatively by visual
comparison of a predicted profile $\sigma_{\rm RMS}(R)$ to the profile
that was inferred from the data and is shown in
Figure~\ref{f:kinprofile}. However, quantitative data-model comparison
is best performed using a maximum likelihood approach. We assume that
the predicted velocity distribution at projected radius $R$ is a
Gaussian with dispersion $\sigma_{\rm RMS}(R)$. This is a reasonable
assumption, at least for isotropic models (e.g., van der Marel \&
Franx 1993). Calculation of the full line-of-sight velocity
distribution shape for spherical anisotropic models is not
straightforward (e.g., Gerhard 1993; van der Marel \etal 2000) and is
outside of the scope of the present paper (see
Section~\ref{ss:anisotropic} for a further discussion of anisotropic
models). Let a star, number $i$, be observed at projected radius
$R_i$. Let the uncertainty in its observed velocity $v_i$ be $\Delta
v_i$. The probability distribution $P_i(v)$ for the velocity $v$
observed for star $i$ is then a Gaussian with dispersion $[\sigma_{\rm
RMS}^2(R_i) + \Delta v_i^2]^{1/2}$, centered on the cluster systemic
velocity $v_{\rm sys}$. The total likelihood $L$ of the dataset is the
product of the quantities $P_i(v_i)$ for all of the $N = 1797$ stars
in the combined velocity sample, so that
\begin{equation}
\label{like}
  \lambda \equiv -2 \ln L =
     \sum_{i=1}^N \ln 2 \pi [\sigma_{\rm RMS}^2(R_i) + \Delta v_i^2] +
     \sum_{i=1}^N (v_i - v_{\rm sys})^2 / 
                  [\sigma_{\rm RMS}^2(R_i) + \Delta v_i^2] .
\end{equation}
The models are generally characterized by a certain set of parameters.
The parameters that yield the minimum $\lambda$ value, $\lambda_{\rm
min}$, are the ones that provide the best fit.

Once a best-fit model has been identified, there are two further
questions that need to be answered: (a) is the best-fitting model
statistically acceptable; and (b) what are the confidence regions
around the best-fitting model parameters? To address the first
question, we note that the last term in equation~(\ref{like}) is a
$\chi^2$ quantity. If each $v_i$ is drawn from a Gaussian distribution
with dispersion $[\sigma_{\rm RMS}^2(R_i) + \Delta v_i^2]^{1/2}$,
which is what we assume here, then this term follows a $\chi^2$
probability distribution with $N$ degrees of freedom. For large $N$,
this distribution has a mean $N$ and dispersion $\sqrt{2N}$. So for
the best-fit model to be statistically acceptable at the $1\sigma$
level, the value $\lambda_{\rm min}$ should be within $\pm \sqrt{2N}$
from the expectation value
\begin{equation}
\label{likerange}
  \langle \lambda \rangle = 
     \sum_{i=1}^N \ln 2 \pi [\sigma_{\rm RMS}^2(R_i) + \Delta v_i^2] + N ,
\end{equation}
where $\sigma_{\rm RMS}(R)$ is the profile predicted by the best-fit
model. To address the second question, we use a well-known theorem of
mathematical statistics (e.g., Stuart \& Ord 1991; used also by
Merritt \& Saha 1993) which states that the likelihood-ratio statistic
$\lambda - \lambda_{\rm min}$ tends to a $\chi^2$ statistic in the
limit of large $N$, with the number of degrees-of-freedom equal to the
number of free parameters that have not yet been varied and chosen so
as to optimize the fit. This is a consequence of the central limit
theorem. Hence, the likelihood-ratio statistic $\lambda - \lambda_{\rm
min}$ reduces to the well-known $\Delta \chi^2$ statistic (e.g., Press
\etal 1992). This allows straightforward calculation of confidence
regions around the best-fitting model parameters.

\subsection{Isotropic Models with Constant Mass-to-Light Ratio}
\label{ss:isoconstant}

We started by calculating isotropic models with a ($V$-band)
mass-to-light ratio $\Upsilon$ that is independent of radius. We allow
for the possibility of a central black hole of mass $M_{\rm BH}$.
Figure~\ref{f:models}a shows the contours of the likelihood quantity
$\lambda$ in the $(\Upsilon,M_{\rm BH})$ plane. The solid contours
show the $1$, $2$, $3$, and $4\sigma$ confidence regions on the
best-fitting model parameters. The best-fitting mass-to-light ratio is
almost independent of $M_{\rm BH}$. This is because the black hole
only influences the dynamical predictions close to the center, so that
$\Upsilon$ is determined more-or-less uniquely by the data at large
radii. By marginalizing over $M_{\rm BH}$, we infer a best-fitting
mass-to-light ratio $\Upsilon = 1.6 \pm 0.1$. For understanding the
dynamical structure of M15, the more interesting quantity is the black
hole mass. By marginalizing over $\Upsilon$ we infer a best-fitting
value $M_{\rm BH} = (3.2 \pm 2.2) \times 10^3 \Msun$. The best-fitting
model has likelihood $\lambda = 13405$. Equation~(\ref{likerange})
predicts that $\lambda$ should be in the range $13419 \pm 60$, so the
best-fit model is statistically acceptable.

Figure~\ref{f:models}b compares the predicted and observed profiles of
$\sigma_{\rm RMS}$ for fixed $\Upsilon = 1.6$ and different black hole
masses. The model with $M_{\rm BH} = 0$ predicts a decrease in
$\sigma_{\rm RMS}$ towards the center. This is a direct consequence of
the fact that the luminosity and number density of M15 increase
steeply towards the center (e.g., Tremaine \etal 1994). The observed
$\sigma_{\rm RMS}$ profile does not show a decrease towards the
center, requiring the presence of a central black hole for the models
to fit the data.

As discussed in Section~\ref{s:intro}, there are independent
constraints on the central mass distribution of M15 from observations
of pulsar accelerations. The measured accelerations of two pulsars
at $R = 1.1''$ from the cluster center yield a lower limit $M_{\rm
cyl,min}(1.1'')$ on the M15 mass $M_{\rm cyl}(1.1'')$ enclosed within
a cylinder of this radius. Equation (3.6) of Phinney (1993), combined
with the pulsar data in his table~2, yields $M_{\rm min}(1.1'') = 4.5
\times 10^3 \Msun$. The most likely (expectation) value for the enclosed mass,
$\langle M_{\rm cyl}(1.1'') \rangle$, depends on the phase-space
distribution of the pulsars in the cluster. Phinney (1993) quotes that
for typical distributions $M_{\rm cyl,min} / \langle M_{\rm cyl}
\rangle = 0.6$--$0.75$. This yields $\langle M_{\rm cyl}(1.1'')
\rangle \approx (6.0$--$7.5) \times 10^3 \Msun$. For the best fit model in
Figure~\ref{f:models} we find that $M_{\rm cyl}(1.1'') = 6.3 \times
10^3 \Msun$ (of which $3.2 \times 10^3 \Msun$ in the form of a black
hole, and $3.1 \times 10^3 \Msun$ in the form of stellar
objects). This is in excellent agreement with the pulsar constraints.

In our stellar dynamical analysis we have used the position of the
cluster center given in Paper~I, which is based on the analysis of
Guhathakurta \etal (1996). This position has a 1-$\sigma$ uncertainty
of $0.2''$ in each coordinate, which corresponds to a two-dimensional
1-$\sigma$ uncertainty of $0.3''$. This is small compared to the
relevant scales in the analysis. For example, the kinematical profiles
in Figure~\ref{f:kinprofile} only start at $R=0.5''$ (this is the
average radius of the innermost 11 data points, where 11 is the adopted
data window size in {\tt SROTATE}). More importantly,
Figure~\ref{f:models}b shows that models without a black hole fail to
fit the data over the entire region $R \lta 6''$. Shifts in the
assumed position of the M15 center of order $\sim 0.3''$ cannot change
this. To quantify this, we repeated the likelihood analysis with
several different assumed positions for the M15 center. These
positions were obtained by shifting the nominal center position by
$\sim 0.3''$ in various directions. The resulting changes in the
best-fitting $M_{\rm BH}$ were found to be $\lta 0.1 \times 10^3
\Msun$. This is much smaller than the formal error on $M_{\rm BH}$
($2.2 \times 10^3 \Msun$) and is therefore negligible. The same
applies to the other models that are discussed in the remainder of
this section. Hence, none of the main results of the present study are
affected by the uncertainties in the exact position of the M15 center.

\subsection{Isotropic Models with Varying Mass-to-Light Ratio}
\label{ss:isovarying}

The assumption of a mass-to-light ratio $\Upsilon$ that does not vary
with radius is obviously a considerable simplification. M15 does not
have a constant density core, which suggests that core-collapse has
occurred as a result of two-body relaxation. A natural consequence of
two-body relaxation is mass segregation. In an attempt to reach
equipartition of energy, heavy stars and dark remnants sink to the
center of the cluster, which causes a central increase in
$\Upsilon(r)$. This boosts the predicted velocities close to the
center, which may obviate the need to invoke a central black hole in
the models (Peterson \etal 1989).

Detailed Fokker-Planck calculations are required to determine whether
mass-segregation resulting from realistic initial conditions can fit
the kinematical data for M15. We have not calculated such models
ourselves. Instead, we have relied on the results published by Dull
\etal (1997), which are the most recently published Fokker-Planck
models for M15. The Dull \etal models were constructed to fit
kinematical data at radii $R \gta 4''$. It is therefore natural to ask
how these models compare to the new data that is now available for $R
\lta 4''$. Dull \etal do not show the kinematical predictions of their
models inside $R = 3''$. However, they do show the run of
$\Upsilon(r)$ in their best-fit model (their figure 12). We measured
this profile from their paper, and used it as input into our Jeans
models. We adopted the profile for the situation that is intermediate
between a collapsed core and an expanded core, which provides the
steepest central increase in $\Upsilon$. We scaled $\Upsilon(r)$
uniformly downward by a factor 0.86, so as to best match the
kinematical predictions shown by Dull \etal at radii $R \gta
3''$. This scaling allows for differences in the adopted distance,
foreground absorption, and surface brightness profile. The resulting
mass-to-light ratio profile does not vary much at the radii $R \gta
0.5''$ for which kinematical data is available, and is between 1.4 and
2.5 throughout. However, at smaller radii $\Upsilon$ increases to
values $\gta 10$, due to a central cusp that is strongly dominated by
non-luminous neutron stars.

We constructed hydrostatic equilibrium models with the $\Upsilon(r)$
profile thus obtained from Dull et al.\footnote{Figure~12 of Dull
\etal (1997) shows the mass-to-light ratio profile after projection
along the line-of-sight, and not the intrinsic three-dimensional
mass-to-light ratio profile. We properly accounted for this in our
modeling.} and with a range of black hole
masses. Figure~\ref{f:maxlike}b compares the predictions for
$\sigma_{\rm RMS}$ to the observed profile from
Figure~\ref{f:kinprofile}d.  Figure~\ref{f:maxlike}a shows the
likelihood quantity $\lambda$ as a function of $M_{\rm BH}$, together
with the $1$ and $2\sigma$ confidence levels. The best-fitting model
has likelihood $\lambda = 13412$. Equation~(\ref{likerange}) predicts
that $\lambda$ should be in the range $13417 \pm 60$, so the best-fit
model is statistically acceptable. The best-fitting black hole mass is
$M_{\rm BH} = (4.5 \pm 2.1) \times 10^3 \Msun$. So even with a
realistic mass-to-light ratio profile $\Upsilon(r)$ inferred from
Fokker-Planck models, the models still require a black hole to fit the
data. The black hole mass is not very different from the value derived
in Section~\ref{ss:isoconstant} under the assumption of a constant
$\Upsilon$. The reason for this is that the enclosed mass inside a
(three-dimensional) radius of $\sim 0.5''$ is only $\sim 1.0 \times
10^3 \Msun$ larger for the Fokker-Planck models than it is for the
constant $\Upsilon$ models. So even though in the former models there
is a central concentration of dark remnants, the total mass in these
remnants is insufficient to explain the observed kinematics. In fact,
with the $\Upsilon(r)$ profile from Dull \etal (1997), the models
require a larger BH mass than for a constant $\Upsilon$ (namely, $[4.5
\pm 2.1] \times 10^3 \Msun$ versus $[3.2 \pm 2.2] \times 10^3
\Msun$). This is because $\Upsilon(r)$ actually decreases radially 
inwards between $\sim 1$ arcmin and $\sim 3''$ (presumably caused by
mass segregation concentrating the giants in the center compared to
less massive stars). The best-fit model in Figure~\ref{f:maxlike} has
a mass $M_{\rm cyl}(1.1'') = 9.1 \times 10^3 \Msun$ within a cylinder
of radius $R = 1.1''$ (of which $4.5 \times 10^3 \Msun$ in the form of
a black hole, and $4.6 \times 10^3 \Msun$ in the form of stellar
objects). This is somewhat larger than the expectation value implied
by the pulsar constraints, $\langle M_{\rm cyl}(1.1'')
\rangle \approx (6.0$--$7.5) \times 10^3 \Msun$ 
(Section~\ref{ss:isoconstant}), but the residual is not statistically
unacceptable.

A large uncertainty in Fokker-Planck models comes from the assumptions
used for the retention factor of neutron stars. Dull
\etal (1997) assume that all neutron stars are retained. However, the
distribution of pulsar kick velocities suggests that most of the
neutron stars (assuming that all pulsars come from neutron stars)
should have been ejected from the cluster due to the low escape
velocity of M15. The best estimates for the retention factor are
generally around a few percent and never go above 10\% (Drukier 1996;
Davies \& Hansen 1998; Pfahl, Rappaport \& Podsiadlowski 2002). Even
more extreme, Hansen \& Phinney (1997) suggest less than 1\% of single
neutron stars would be retained.  Binaries, however, provide a
mechanism in which the neutron star retention factor may be increased,
but, even when assuming a large binary fraction, the retention factor
is still below 10\% (Davies \& Hansen 1998; Pfahl, Rappaport \&
Podsiadlowski 2002). Thus, it appears that the contribution from
neutron stars may not be very significant for the central mass
distribution in globular clusters. Heavy white dwarfs (1--$1.3 \Msun$)
do not suffer from these large birth kicks and so can provide some
contribution to the central mass, but these objects are not as
centrally-concentrated (Dull \etal 1997). The best way to understand
the effect of stellar remnants is to include evolutionary models with
realistic assumptions for the retention factor. The important
conclusion in the present context is that the Dull \etal models
strongly overestimate the expected neutron star retention, yet still
fail to fit the M15 data without a central black hole.

The number of stars that end up as neutron stars or white dwarfs in
Fokker-Planck models depends critically on the assumed initial mass
function (IMF). The more stars there are at the high end of the IMF,
the higher the predicted central velocity dispersion at the present
epoch (e.g., Grabhorn \etal 1992). However, there are good
observational constraints on the present-day local mass function (MF)
of M15, at least below the main-sequence turn-off ($\sim 0.7 \Msun$
for M15). The IMF can therefore not be treated as a completely free
parameter. The MF obtained from HST photometry by De Marchi \& Paresce
(1995) for a field at $R = 4.6$ arcmin from the M15 center has a
power-law slope $x \approx 1.3$ at masses above $\sim 0.25 \Msun$
(where $x$ is defined so that a Salpeter MF has $x = 1.35$). Dull
\etal (1997) use a global IMF slope $x=0.9$ at these masses, which yields 
a local present-day MF slope $x \approx 1.2$ at $R = 4.6$ arcmin
(figure~8 of Grabhorn \etal 1992). This indicates that the Dull \etal
IMF is approximately correct, and if anything, is slightly too
shallow. So there is little room for an increased number of high mass
stars in the IMF, which would be needed to bring the Dull \etal models
in better agreement with the kinematical data presented here. One
would actually like the Fokker-Planck models to reproduce also the
observational constraints on the radial dependence of the present-day
MF (De Marchi \& Paresce 1995; Sosin \& King 1997). A preliminary
report of attempts at this (Murphy \etal 1997) indicates that such
models predict a present-day total number of $\sim 7000$ non-luminous
$1.4 \Msun$ objects in M15. This is 30\% less than the number of such
objects in the Dull \etal models. Hence, more accurate modeling of the
IMF is not likely to improve the fit to the kinematical data presented
here.

The Fokker-Planck models of Dull \etal (1997) are the most recent ones
to have been constructed for M15, but they are not the only ones. The
Dull \etal models built on previous work by Grabhorn \etal
(1992). Both papers used the same Fokker-Planck methodology and
adopted the same IMF for their best fit. The main difference is that
Dull \etal had access to more modern and extensive data to compare
their models to. The models of Dull \etal therefore supersede the
models of Grabhorn et al. Also, the discussion that was
presented above with respect to the Dull \etal models applies equally
to the models of Grabhorn et al. In another study, Phinney (1993)
fitted an independent set of Fokker-Planck models to M15 data. These
models were calculated by Bryan Murphy using an extension of the code
described by Murphy, Cohn \& Hut (1990). These models had several nice
features, including explicit inclusion of stellar evolution and
neutron-star escape.  However, the models used a pure power-law
IMF. This is less accurate than the IMF used by Grabhorn \etal and
Dull \etal and is not consistent with our current understanding of
globular cluster IMFs (e.g., Paresce \& De Marchi 2000). As mentioned
above, a model cannot be considered to be acceptable unless it fits
observational constraints on the present-day MF. Also, Phinney's model
that comes closest to fitting the presently available kinematical data
for M15 (his model z120) must be evolved for $21.4 \,{\rm Gyr}$. This
considerably exceeds the accepted age of the Universe (e.g.,
Lineweaver 1999).

The above arguments indicate that there is an urgent need for more
detailed Fokker-Planck modeling of the structure and evolution of
M15. The models should include as much as possible of the relevant
physics, including neutron star escape. The most current datasets must
be used to constrain the cluster density, kinematics and IMF. With
such a study it will be possible to explore in much more detail than
has been possible here whether the observed kinematics of M15, both
the central $\sigma_{\rm RMS}$ and its dependence on radius, can be
explained without requiring a central black hole.

\subsection{Anisotropic Models}
\label{ss:anisotropic}

To further explore the range of models that can reproduce the observed
kinematical profiles for M15 we have also constructed models with an
anisotropic velocity distribution. Models with radial velocity
dispersion anisotropy predict a larger projected velocity dispersion
near the center of a stellar system than models with an isotropic
velocity dispersion (e.g., Binney \& Mamon 1982). Radially anisotropic
models for M15 may therefore be able to fit the data without invoking
a central black hole. To test how much radial anisotropy is required
for this, we constructed models in which the anisotropy function
$\beta \equiv 1 - \sigma_t^2 / \sigma_r^2$ is parameterized as
\begin{equation}
  \label{betaparam}
  \beta(r) = \beta_0 + (\beta_\infty - \beta_0) \frac{r^2}{r^2 + a^2} .
\end{equation}
This parameterization yields a convenient semi-analytical expression
for the solution of the Jeans equation (van der Marel 1994). It
corresponds to a profile that varies from $\beta = \beta_0$ at the
center to $\beta = \beta_\infty$ at large radii. The transition occurs
at the scale radius $a$.

The value of $\beta_\infty$ has little influence on the dynamical
predictions near the center of the cluster, which is where isotropic
models without a central black hole fail to fit the data
(cf.~Figure~\ref{f:models}). Without loss of generality we therefore
kept $\beta_\infty$ fixed at zero in our study. We also assumed that
the mass-to-light ratio $\Upsilon$ is constant as a function of
radius, and we did not include a black hole (i.e., $M_{\rm BH} = 0$).
This leaves three free parameters, $\beta_0$, $a$ and $\Upsilon$,
which were varied to optimize the fit to the data. The best-fitting
model was found to have $\beta_0 = 0.65$, $a = 18''$ and $\Upsilon =
1.6$. The inferred mass-to-light ratio is similar to that inferred for
isotropic models (Section~\ref{ss:isoconstant}). The best-fitting
model has likelihood $\lambda = 13401$. Equation~(\ref{likerange})
predicts that $\lambda$ should be in the range $13395 \pm 60$, so the
model is statistically acceptable.  Figure~\ref{f:anisotropy}b
compares the predictions for the best-fitting anisotropic model
(dotted curve) to the observed $\sigma_{\rm RMS}$ from
Figure~\ref{f:kinprofile}d. For comparison, we also show the
predictions for models in which $\beta_0$ was kept fixed at $\beta_0 =
0$, $0.2$, $0.4$ and $0.8$, and in which only $a$ and $\Upsilon$ were
varied to optimize the fit. Figure~\ref{f:anisotropy}a shows the
likelihood quantity $\lambda$ for these fits as a function of
$\beta_0$, together with the $1$ and $2\sigma$ confidence levels. The
best fit and its $1\sigma$ error are $\beta_0 = 0.65 \pm 0.2$.

The models in Figure~\ref{f:anisotropy} show that it is possible to
fit the M15 data without invoking a central mass
concentration. However, this requires that $\sigma_r = 1.6 \sigma_t$
(i.e., $\beta \approx 0.65$) throughout much of the central region of
M15. Such a high value is difficult to reconcile with our current
understanding of the dynamical structure of globular clusters. Even
elliptical galaxies, which are collisionless, do not have such large
anisotropies (e.g., Kronawitter \etal 2000). By contrast, in a cluster
like M15 considerable two-body relaxation must have occurred near the
center, which tends to isotropize the velocity distribution (Peterson
\etal 1989). This has been demonstrated explicitly, both with anisotropic
Fokker-Planck calculations in energy--angular momentum space and with
$N$-body calculations (Takahashi 1996; Baumgardt, Heggie \& Hut
2002). In the region that contains the central $\sim 20$\% of the
cluster mass, the anisotropy parameter $\beta$ remains between $0$ and
$0.1$ at all times during the cluster evolution. Larger radial
anisotropies can develop in the outskirts of the cluster, but we
verified explicitly that this does not change the interpretation of the
central kinematics of M15.

\section{Discussion and Conclusions}
\label{s:conc}

We have obtained high spatial resolution spectroscopy of the central
region of the globular cluster M15 with the STIS spectrograph on board
HST. The observational setup, calibration and spectral extraction were
discussed in Paper~I. Here we have analyzed the spectra with a
cross-correlation technique to determine the line-of-sight velocities
of individual stars. Our final STIS velocity sample contains 64 stars.
Two-thirds of the stars in this sample have their velocity measured
for the first time. Half of the stars reside within a projected radius
$R = 2.4''$ from the center of M15. The new data set triples the
number of stars with measured velocities in the central $R \leq 1''$
of M15 and doubles the number in the central $R \leq 2''$. Our
analysis includes the necessary (small) corrections for the effects of
blending with neighboring stars. Detailed tests on a calibration star
and comparison to ground-based M15 data demonstrate that our velocities
are accurate and trustworthy. We combined the STIS results with
existing ground-based data to obtain a total sample of 1797 stars in
M15 with known line-of-sight velocities. We use the combined sample to
determine the radial profiles of the most important projected
kinematical quantities: the rotation velocity $V_{\rm rot}$; the
position angle of the kinematical major axis, ${\rm PA}_{\rm kin}$;
the velocity dispersion, $\sigma$; and the RMS velocity averaged over
rings on the projected plane of the sky, $\sigma_{\rm RMS}$. Our
results differ from earlier work only in the central few arcsec. In
particular, we find that $\sigma_{\rm RMS}$ rises to $\sim 14 \kms$ at
the innermost radii. This is somewhat higher than the value of
$10$--$12 \kms$ inferred previously from ground-based data (Dull \etal
1997; Gebhardt \etal 2000a).

To interpret the results we constructed dynamical models based on the
Jeans equation for a spherical system. We compared the model
predictions to the data using a maximum-likelihood technique to obtain
the best-fitting model parameters and their confidence regions. If the
velocity distribution is isotropic, then M15 must have a central
concentration of non-luminous material. This could be due to an
intermediate-mass black hole. If one were reluctant to invoke such an
object, then one alternative may be that M15 has a central collection
of dark remnants (neutron stars and/or stellar mass black holes). This
arises naturally in a globular cluster due to the mass segregation
that occurs as two-body relaxation drives the system to
equipartition. However, we argued that the best-fitting Fokker-Planck
models that have previously been constructed for M15 (Dull \etal 1997)
do not predict a large enough concentration of dark remnants to fit
the data. It remains to be seen whether alternative Fokker-Planck
models can be constructed that generate a more massive concentration
of dark remnants from plausible initial conditions. It is useful to
note in this context that an important uncertainty in Fokker-Planck
models comes from the assumptions used for the retention factor of
neutron stars. Dull \etal assumed that all neutron stars are retained,
in contrast with most recent work which predicts that only $\lta 10$\%
will be retained. So Dull \etal may actually have overestimated the
central concentration of dark remnants in their models. Another
alternative scenario is to assume that deviations from isotropy in the
velocity distribution may be responsible for the observed
kinematics. However, to fit the kinematical data without any mass
concentration one must assume that the velocity distribution is
radially anisotropic near the center, $\beta_0 = 0.65 \pm 0.2$. This
contradicts the predictions of both Fokker-Planck models and $N$-body
calculations, which suggest that the velocity distribution in the
central region of a globular clusters remains close to isotropic at
all times during the cluster evolution.

In view of the results that we have presented, the presence of an
intermediate mass black hole in M15 appears to be the most plausible
explanation of the data. As noted in Section~\ref{s:intro}, there are
several mechanisms by which such a black hole could plausibly have
formed. For the best-fit black hole mass we adopt the average of the
values which were inferred in Sections~\ref{ss:isoconstant}
and~\ref{ss:isovarying} using a constant mass-to-light ratio and the
Dull \etal (1997) mass-to-light ratio profile, respectively. This
yields: $M_{\rm BH} = (3.9 \pm 2.2) \times 10^3 \Msun$. This mass is
consistent with the constraints on the central mass distribution of
M15 implied by observations of pulsar accelerations (Phinney 1993).
The black hole mass inferred for M15 matches remarkably well with the
understanding that has been developed for the presence of black holes
in the centers of galaxies. For these black holes, there is a strong
correlation between the black hole mass and the velocity dispersion of
the bulge component (Gebhardt \etal 2000b; Ferrarese \& Merritt 2000).
Figure~\ref{f:bhcorr} shows the available data points and the best fit
from the recent compilation of Tremaine \etal (2002). For M15, the
luminosity weighted mean velocity dispersion within the half light
radius ($1.06$ arcmin, Harris 1996) is $12.1 \kms$ (this quantity was
defined and calculated similarly as in Gebhardt \etal 2000b). At this
dispersion, the estimated black hole mass fits perfectly on the
extrapolation of the relation established for galaxies
(Figure~\ref{f:bhcorr}).  Interestingly, a study of the globular
cluster G1 in the Andromeda galaxy, performed simultaneously with the
present study, has also provided evidence for a central black hole
(Gebhardt, Rich \& Ho 2002). Like M15, this globular cluster fits
perfectly on the relation shown in Figure~\ref{f:bhcorr}. This
independent research strengthens the interpretation of the M15 data in
terms of an intermediate-mass black hole. It has generally been
believed that globular clusters and galaxies form and evolve quite
differently, so it could be that it is a mere coincidence that they
fall on the same $M_{\rm bh}$---$\sigma$ relationship. However, it
could also have some deep physical significance. For example, it may
point to a new link between galaxy formation and globular cluster
formation. Or it may point to a link between the black holes in these
systems. For example, it could be that the massive black holes in
galaxies grew from seed black holes that arose in clusters. There may
also be a link with the intermediate luminosity X-ray objects that are
known to exist in external galaxies, and which have been argued to be
intermediate mass black holes. These issues will need to be explored
with future observational and theoretical studies.

Despite the interesting evidence for the presence of an intermediate
mass black hole in M15, some words of caution are justified. All of
the dynamical models that have been constructed for M15 remain
somewhat idealized. This is true both for the Jeans models presented
here and for the Fokker-Planck models presented elsewhere. For
example, the Jeans models assume exact hydrostatic equilibrium, which
is generally expected to be a good assumption (see
Section~\ref{ss:modtechnique}). Nonetheless, during periods of
particularly rapid evolution in the cluster structure this assumption
could yield results that are biased. Fokker-Planck models can address
the cluster evolution directly and do not need to rely on the
assumption of hydrostatic equilibrium. On the other hand, the results
of Fokker-Planck models depend strongly on the processes of stellar
evolution and binary heating, both of which are generally modeled only
in rudimentary ways. More generally, globular clusters are complicated
systems from a theoretical viewpoint, much more so than galaxies, and
not all of the essential physics may yet have been fully
understood. There is some evidence from observations that this may
indeed be the case. For example, M15 rotates quite rapidly in the
central regions (cf.~Figure~\ref{f:kinprofile}), and this is not
naturally explained by any theoretical model (Gebhardt
\etal 2000a). Also, the $\sigma_{\rm RMS}$ of M15 appears to have a
small dip at intermediate radii (at $R \approx 13''$), which is not
naturally explained by any of the models constructed here
(Figures~\ref{f:models}--\ref{f:anisotropy}). Further studies to test
the observational reality of these features would be valuable, as
would further theoretical work to address their origin.

We have assumed throughout our study, as has previous work on M15,
that the observed kinematics are characteristic of the cluster, and
are not contaminated by possible orbital motion of stars in binary
systems. There are several reasons that make this a reasonable
assumption (Hut \etal 1992). The binary fraction of globular clusters
is believed to be only of order 10\%. Also, most of the stars in the
velocity sample are red giants. Their relatively large radii imply
that any binaries must have large separations and orbital velocities
$\lta 25 \kms$. Inclination, phase and ellipticity effects imply that
for an average binary at a random epoch only a fraction of the
velocity amplitude will be observed along the line of sight. These
issues conspire to make it extremely challenging to identify even a
few binaries in globular clusters from large line-of-sight velocity
studies, even with high quality multi-epoch data (Pryor \etal 1989).
One can turn this around to argue that the average observed kinematics
of large samples of stars should not be influenced significantly by
any orbital motion in binaries. This assumption is supported by the
fact that the mass distribution inferred here from stellar kinematics
agrees with that inferred from pulsar studies. Nonetheless, it would
be useful for future studies to attempt a detailed quantitative
assessment of the potential contamination of the line-of-sight
kinematics of globular clusters by binaries.

In the future it may be possible to strengthen the observational
constraints on the central structure of M15 through proper motion
studies. With two additional velocity components it will be possible
to directly establish the (an)isotropy of the stellar velocity
distribution. It has been demonstrated that such studies are feasible
with HST (e.g., Anderson \& King 2000). However, the severe crowding
in the central few arcsec of M15 may provide a significant hurdle to
overcome. An alternative way to strengthen the observational
constraints would be to increase the sample of radial velocities. This
would require inclusion of fainter stars near the turnoff magnitude
($V \approx 19$ in M15), which were inaccessible to our STIS study
because of limited $S/N$. In principle, a 4m class telescope has
sufficient light gathering power to perform such a study in a
reasonable amount of time. By centering the spectra around the CO
band-head ($2.3{\mu}{\rm m}$), it may be possible to take full
advantage of adaptive optics to attain a spatial resolution comparable
to that of HST.

%%%%%%%%%%%%%%%
% Acknowledgments
%%%%%%%%%%%%%%%

\acknowledgments
Support for proposals \#8262 was provided by NASA through a grant from
the Space Telescope Science Institute, which is operated by the
Association of Universities for Research in Astronomy, Inc., under
NASA contract NAS 5-26555. We thank Pierre Dubath for helpful advice
in the early stages of this project. We thank the anonymous referee
for useful feedback that helped improve the presentation of the paper.

\clearpage

%%%%%%%%%%%%%%%
% Appendix, if any
%%%%%%%%%%%%%%%

% \appendix

% \section{}
% \label{s:AppA}

%%%%%%%%%%%%%%%
% Use a small baselineskip for the references, unless in submission mode.
%%%%%%%%%%%%%%%

\ifsubmode\else
\baselineskip=10pt
\fi

%%%%%%%%%%%%%%%
% Reference List
%%%%%%%%%%%%%%%

\clearpage

%%%%%%%%%%%%%%%
% Change back to the regular baselineskip, if necessary
%%%%%%%%%%%%%%%

\ifsubmode\else
\baselineskip=14pt
\fi

%%%%%%%%%%%%%%%
% Figure Captions
%%%%%%%%%%%%%%%

\newcommand{\figcapcalibconf}{Radial velocity determination tests for
294 HST/STIS spectra of the calibration star. {\bf (a; left panel)}
Absolute difference $|v_{\rm diff}|$ (in $\kms$) between the measured
radial velocity and true calibration star velocity as a function of
the cross-correlation statistic $r_{\rm cc}$. Open symbols indicate
spectra with average signal-to-noise ratio $S/N \geq 5.5$ per pixel
and closed symbols indicate spectra with $S/N < 5.5$. The vertical
line indicates the cutoff $r_{\rm cc} = 2$ that we have used in the
analysis. {\bf (b; right panel)} Absolute difference $|v_{\rm diff}|$
(in $\kms$) between the measured radial velocity and true calibration
star velocity as a function of $S/N$. Open symbols indicate spectra
with $r_{\rm cc} \geq 2$ and closed symbols indicate spectra with
$r_{\rm cc} < 2$. The vertical line indicates the cutoff $S/N = 5.5$
that we have used in the analysis. Both panels show a pronounced
transition from unreliable measurements to reliable measurements,
either with increasing $r_{\rm cc}$ or with increasing $S/N$.  Spectra
that have both $r_{\rm cc} \geq 2$ and $S/N \geq 5.5$ (the open
symbols to the right of the cut-off line in each panel) yield a
reliable measurement of the stellar velocity. The horizontal line in
both panels corresponds to $2.5 \kms$, which is the systematic
uncertainty in the velocity calibration of the data (cf.~Paper~I).
The axes in both panels are logarithmic.\label{f:calibconf}}

\newcommand{\figcapclusterconf}{Cross-correlation results for
HST/STIS spectra of M15 stars. Spectra with an average signal-to-noise
ratio $S/N \geq 5.5$ per pixel were extracted for 131 stars. The
ordinate of the figure shows the absolute difference $|v_{\rm diff}|$
of the inferred stellar velocity and the systemic velocity of M15,
divided by the central velocity dispersion of M15 (assumed to be $12
\kms$, Gebhardt \etal 2000). This `normalized velocity deviation' is shown 
as function of cross-correlation statistic $r_{\rm cc}$. A large value
of the normalized velocity deviation generally indicates that the
inferred stellar velocity is incorrect. There is a pronounced
transition from unreliable measurements to reliable measurements with
increasing $r_{\rm cc}$.  The cutoff at $r_{\rm cc} = 2$ that we have
used in our analysis (vertical line) excludes all measurements that
are obviously unreliable. This confirms the analysis of the
calibration star spectra presented in
Figure~\ref{f:calibconf}.\label{f:clusterconf}}

\newcommand{\figcapcmd}{Color-magnitude diagram of the stars in M15,
obtained from the photometric HST/WFPC2 imaging catalog presented in
Paper~I. The crosses indicate the location of the stars for which we
have derived reliable radial velocities from the HST/STIS spectra.
The two encircled crosses indicate potential RR Lyrae variables which
were not included in the dynamical modeling.\label{f:cmd}}

\newcommand{\figcapdisthist}{Distribution of the number of M15 stars $N$ 
with known line-of-sight velocities (with errors $\leq 10 \kms$) as
function of projected radius from the cluster center. {\bf (a; left
panel)} Binned histogram. Solid line: the HST/STIS velocity sample;
Dashed line: the sample with ground-based measurements compiled by
Gebhardt \etal (2000a). {\bf (b; right panel)} Logarithm of the
cumulative distribution (i.e., total number of stars within a given
radius). Solid line: the HST/STIS velocity sample; Dashed line: the
Gebhardt \etal (2000a) compilation; Heavy solid line: the combined HST/STIS
and ground-based sample, corrected for double entries (i.e., stars
present in both samples are counted only once). The HST/STIS data
significantly increase the number of stars with known line-of-sight
velocities close to the cluster center.\label{f:disthist}}

\newcommand{\figcapxyplot}{Positions of the stars with known
line-of-sight velocities in the central $4 \times 4$ arcsec of M15.
Coordinates are (RA, DEC) with respect to the cluster center
determined by Guhathakurta \etal (1996).  Filled dots: stars with a
velocity measurement from STIS; Circles: stars with a ground-based
velocity measurement. The symbol size provides a measure of the
brightness of the star; the circles were increased in size relative to
the filled dots to avoid overlapping symbols. The STIS measurements
cluster around a line with a position angle of $26.65^{\circ}$, which
was the slit position angle used for the observations (see
Paper~I).\label{f:xyplot}}

\newcommand{\figcapvelcomp}{Comparison of the velocities derived with
HST/STIS to those obtained from ground-based data, for those stars for
which both exist. The excellent agreement shows that there are no
unidentified systematic errors in the data. The two labeled stars are
discussed in the text.\label{f:velcomp}}

\newcommand{\figcapresiduals}{Velocity residuals, $v_{\rm HST} -
v_{\rm ground}$, for those stars that have both HST/STIS and
ground-based measurements of their line-of-sight velocities. {\bf (a;
top panel)} Residuals as a function of stellar magnitude $V$.  {\bf
(b; bottom panel)} Residuals as a function of broad-band color $B-V$.
The weighted mean velocity residual is $0.2 \pm 1.4
\kms$.\label{f:residuals}}

\newcommand{\figcaprepeat}{Histogram of the differences $\Delta v_{ij}$ 
between multiple velocity measurements of the same star. The histogram
includes those stars for which multiple independent STIS spectra are
available that individually satisfy the selection criteria with
respect to blending, $S/N$ and $r_{\rm cc}$ discussed in the text. The
histogram is well fit by a Gaussian with a dispersion of $6 \kms$
(solid curve). The stellar velocities themselves can be estimated with
an uncertainty that is factor 2 smaller than this (because the
statistical uncertainty in the average of two measurements is a factor
2 smaller than the statistical uncertainty in their
difference).\label{f:repeat}}

\newcommand{\figcapkinprofile}{Radial profiles of projected kinematical
quantities inferred from the combined HST/STIS and ground-based
stellar line-of-sight velocity samples. {\bf (a; top left panel)}
Velocity dispersion $\sigma$; {\bf (b; top right panel)} Rotation
velocity $V_{\rm rot}$; {\bf (c; bottom left panel)} Position angle
${\rm PA}_{\rm kin}$ of the kinematical major axis; {\bf (d; bottom
right panel)} The RMS projected line-of-sight velocity $\sigma_{\rm
RMS} \equiv \langle V_{\rm rot}^2 + \sigma^2 \rangle^{1/2}$, where the
angle brackets denote the average over a ring on the projected plane
of the sky. Solid curves are best estimates; dotted curves define the
$68.3$\% confidence bands.\label{f:kinprofile}}

\newcommand{\figcaptrend}{Curves show the radial profile of 
$\sigma_{\rm RMS}$ with its associated uncertainty, as in
Figure~\ref{f:kinprofile}d. For comparison, data points show the
absolute value of the difference $\Delta v \equiv v_{*} - v_{\rm sys}$
between the measured velocities of individual stars and the M15
systemic velocity. This allows some visual assessment of the radial
trends in the data. The $\sigma_{\rm RMS}$ curves start at $R=0.5''$,
which is the average radius of the innermost 11 data points (11 is the
adopted data window size in {\tt SROTATE}). The radial distribution of
the data points reflects not only the distribution of stars in M15,
but also observational selection bias.\label{f:trend}}

\newcommand{\figcapmodels}{Data-model comparison for spherical dynamical
models with an isotropic velocity distribution, a constant
mass-to-light ratio $\Upsilon$ (in solar $V$-band units), and a
central black hole of mass $M_{\rm BH}$. {\bf (a; left panel)}
Likelihood contours as function of $\Upsilon$ and $M_{\rm BH}$. The
best-fit model has $\Upsilon = 1.6 \pm 0.1$ and $M_{\rm BH} = (3.2 \pm
2.2) \times 10^3 \Msun$. Solid curves indicate the $1$, $2$, $3\sigma$
and $4\sigma$ confidence regions in the two-dimensional
$(\Upsilon,M_{\rm BH})$ plane. Dotted curves show likelihood contours
inside the $1\sigma$ region. {\bf (b; right panel)} The RMS projected
line-of-sight velocity $\sigma_{\rm RMS}$ as a function of projected
radius $R$. The heavy jagged curve surrounded by heavy dashed curves
is the observed profile, as in Figure~\ref{f:kinprofile}d. The smooth
thin curves are the predictions for models with $\Upsilon = 1.6$ and
$M_{\rm BH}$ ranging from 0 to $10 \times 10^3 \Msun$ in steps of
$10^3 \Msun$. The dotted curve highlights the predictions for $M_{\rm
BH} = 3 \times 10^3 \Msun$, which is closest to the best-fit value
$M_{\rm BH} = (3.2 \pm 2.2) \times 10^3 \Msun$.\label{f:models}}

\newcommand{\figcapmaxlike}{Data-model comparison for spherical dynamical
models with an isotropic velocity distribution, a mass-to-light ratio
profile $\Upsilon(r)$ inferred from Fokker-Planck models (Dull \etal
1997), and a central black hole of mass $M_{\rm BH}$. {\bf (a; left
panel)} The likelihood quantity $\lambda$ defined in
equation~(\ref{like}) as function of $M_{\rm BH}$. The minimum in
$\lambda$ identifies the best fit black hole mass. Horizontal dashed
lines indicate the $1$ and $2\sigma$ confidence regions. {\bf (b;
right panel)} The RMS projected line-of-sight velocity $\sigma_{\rm
RMS}$ as a function of projected radius $R$. The heavy jagged curve
surrounded by heavy dashed curves is the observed profile, as in
Figure~\ref{f:kinprofile}d. The smooth thin curves are the predictions
for models with $M_{\rm BH}$ ranging from 0 to $10
\times 10^3 \Msun$ in steps of $10^3 \Msun$. The dotted curves highlight
the predictions for $M_{\rm BH} = 4 \times 10^3 \Msun$ and $M_{\rm BH}
= 5 \times 10^3 \Msun$, which bracket the best fit value $M_{\rm BH} =
(4.5 \pm 2.1) \times 10^3 \Msun$.\label{f:maxlike}}

\newcommand{\figcapanisotropy}{Data-model comparison for spherical dynamical
models with an anisotropic velocity distribution parameterized by
equation~(\ref{betaparam}), a constant mass-to-light ratio $\Upsilon$,
and no central black hole. {\bf (a; left panel)} The likelihood
quantity $\lambda$ defined in equation~(\ref{like}) as function of the
central anisotropy $\beta_0$. At each $\beta_0$, the parameters $a$
and $\Upsilon$ were varied to optimize the fit; $\beta_\infty$ was
kept fixed at zero. The minimum in $\lambda$ identifies the best fit
$\beta_0$. Horizontal dashed lines indicate the $1$ and $2\sigma$
confidence regions. {\bf (b; right panel)} The RMS projected
line-of-sight velocity $\sigma_{\rm RMS}$ as a function of projected
radius $R$. The heavy jagged curve surrounded by heavy dashed curves
is the observed profile, as in Figure~\ref{f:kinprofile}d. The smooth
thin curves are the predictions for the best-fit models with $\beta_0
= 0$, $0.2$, $0.4$, $0.65$ and $0.8$. The dotted curve is for $\beta_0
= 0.65$, which provides the overall best fit. This indicates that
considerable radial anisotropy must be invoked near the cluster center
to explain the observations without invoking a central mass
concentration.\label{f:anisotropy}}

\newcommand{\figcapbhcorr}{The black hole mass $M_{\rm BH}$ versus
velocity dispersion $\sigma$. Solid points are measurements for
various types of galaxies from the compilation of Tremaine \etal
(2002). The solid line is the best linear fit, and the dashed lines
show the $1\sigma$ confidence band. The black hole mass estimate for
M15 obtained from isotropic models, $M_{\rm BH} = (3.9 \pm 2.2)
\times 10^3 \Msun$ (open point; this is the average of the masses 
inferred in Sections~\ref{ss:isoconstant} and~\ref{ss:isovarying}),
fits perfectly onto the correlation.\label{f:bhcorr}}

%%%%%%%%%%%%%%%
% Figures (in submission mode captions only, unless \printfigtrue)
%%%%%%%%%%%%%%%

\ifsubmode
\figcaption{\figcapcalibconf}
\figcaption{\figcapclusterconf}
\figcaption{\figcapcmd}
\figcaption{\figcapdisthist}
\figcaption{\figcapxyplot}
\figcaption{\figcapvelcomp}
\figcaption{\figcapresiduals}
\figcaption{\figcaprepeat}
\figcaption{\figcapkinprofile}
\figcaption{\figcaptrend}
\figcaption{\figcapmodels}
\figcaption{\figcapmaxlike}
\figcaption{\figcapanisotropy}
\figcaption{\figcapbhcorr}
\clearpage
\else\printfigtrue\fi

\ifprintfig

%%% FIGURE %%%

\clearpage
\begin{figure}
\epsfxsize=0.9\hsize
\centerline{\epsfbox{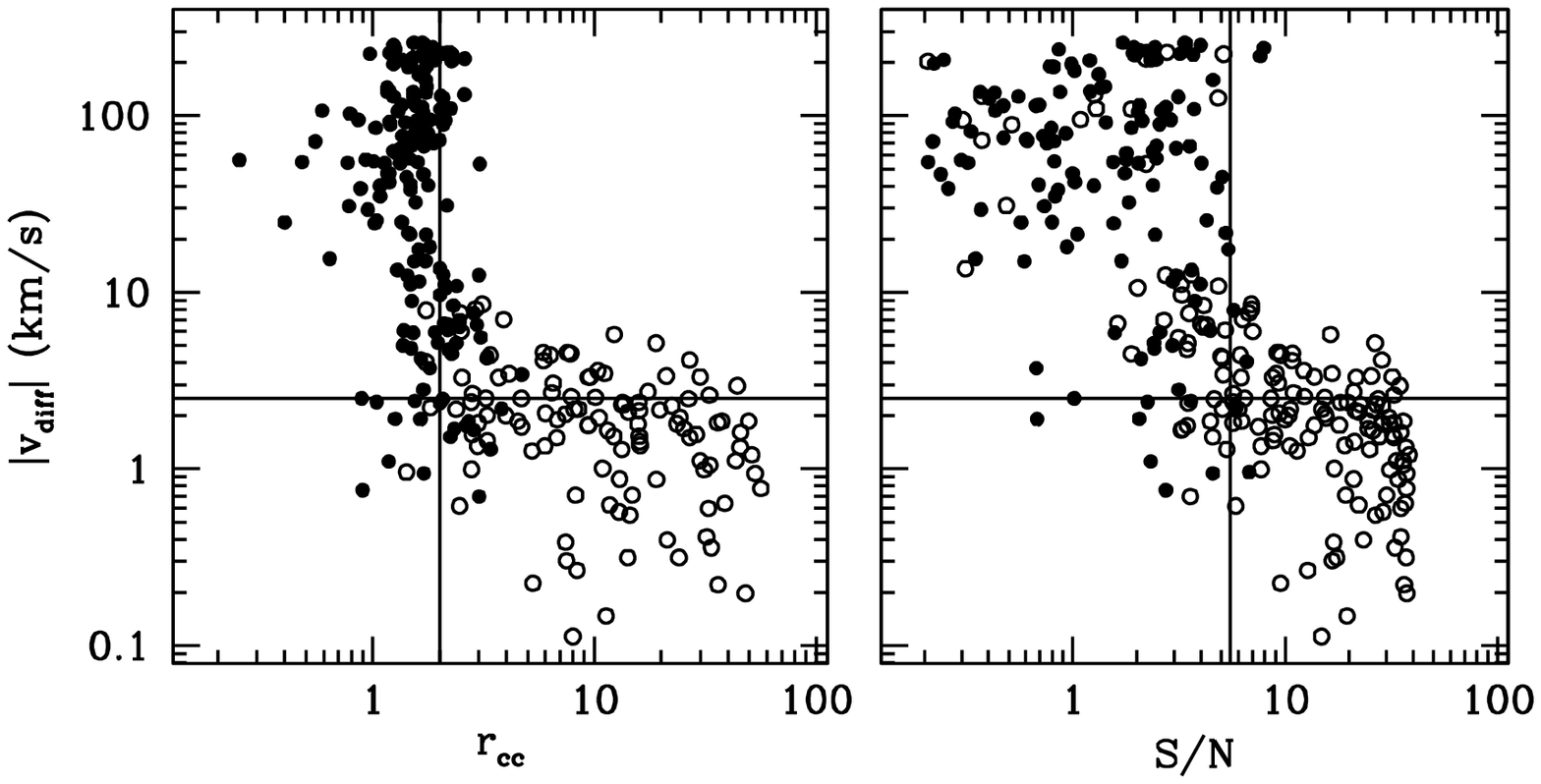}}
\ifsubmode
\vskip3.0truecm
\setcounter{figure}{0}
\addtocounter{figure}{1}
\centerline{Figure~\thefigure}
\else\figcaption{\figcapcalibconf}\fi
\end{figure}

%%% FIGURE %%%

\clearpage
\begin{figure}
\epsfxsize=0.9\hsize
\centerline{\epsfbox{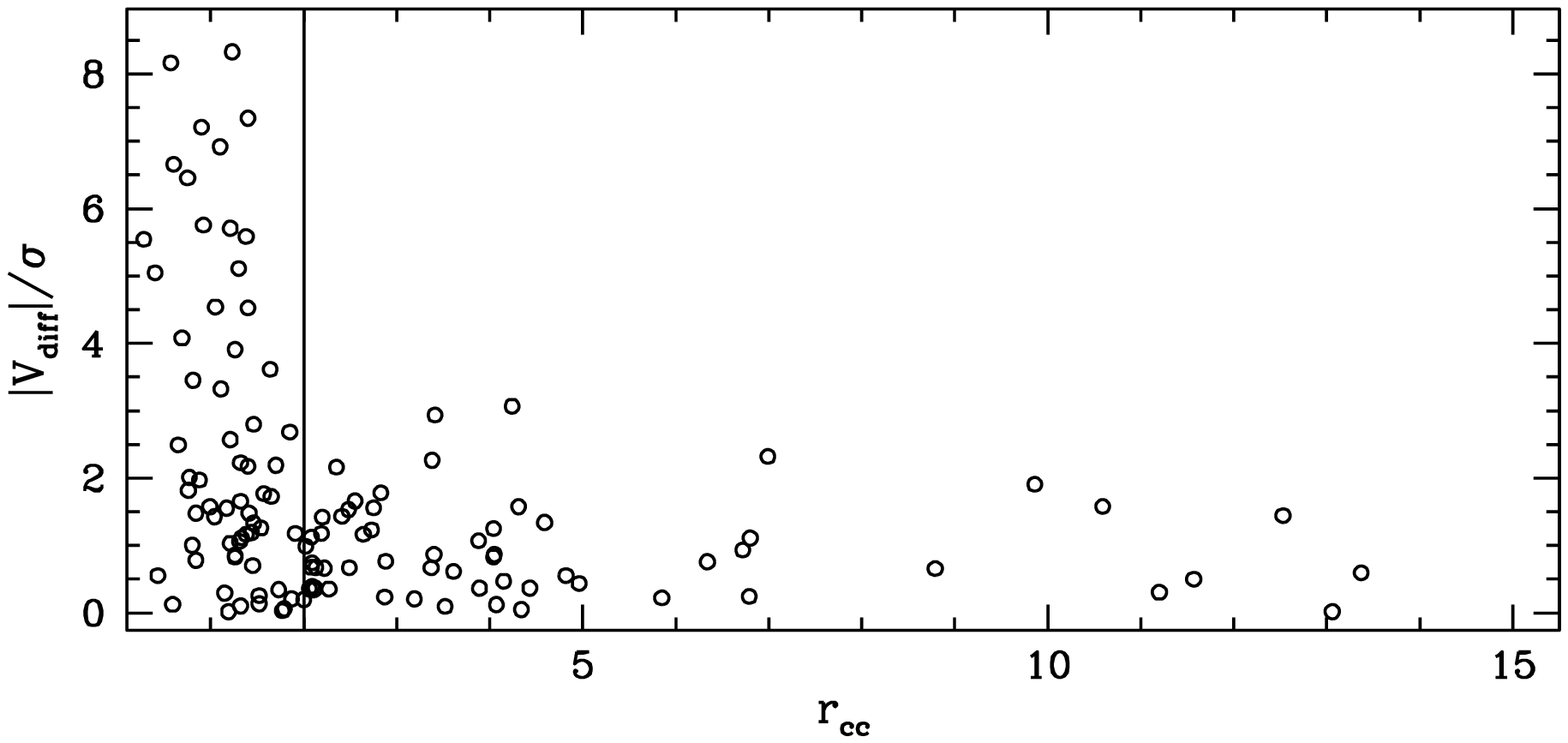}}
\ifsubmode
\vskip3.0truecm
\addtocounter{figure}{1}
\centerline{Figure~\thefigure}
\else\figcaption{\figcapclusterconf}\fi
\end{figure}

%%% FIGURE %%%

\clearpage
\begin{figure}
\epsfxsize=0.9\hsize
\centerline{\epsfbox{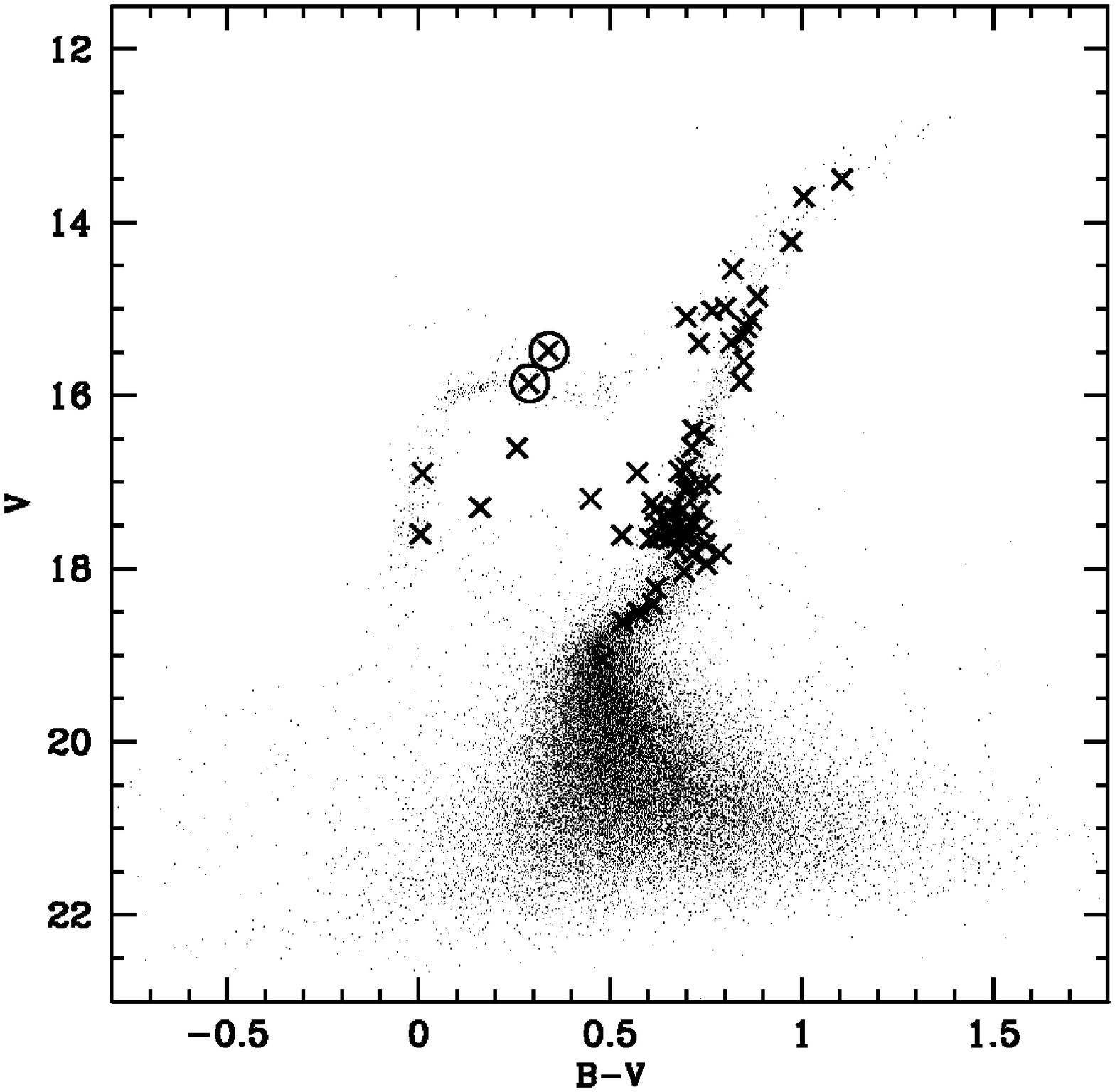}}
\ifsubmode
\vskip3.0truecm
\addtocounter{figure}{1}
\centerline{Figure~\thefigure}
\else\figcaption{\figcapcmd}\fi
\end{figure}

%%% FIGURE %%%

\clearpage
\begin{figure}
\epsfxsize=0.9\hsize
\centerline{\epsfbox{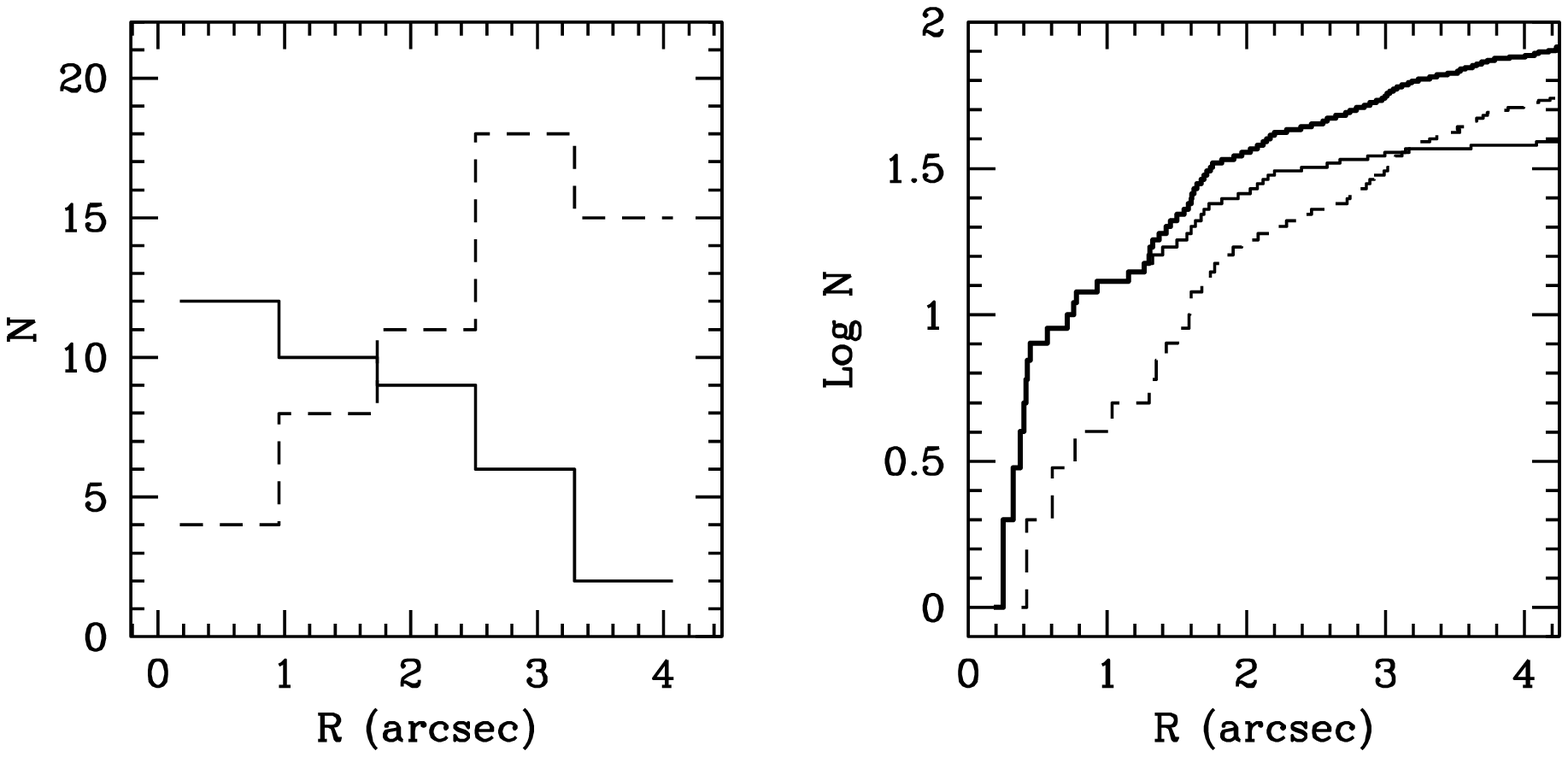}}
\ifsubmode
\vskip3.0truecm
\addtocounter{figure}{1}
\centerline{Figure~\thefigure}
\else\figcaption{\figcapdisthist}\fi
\end{figure}

%%% FIGURE %%%

\clearpage
\begin{figure}
\epsfxsize=0.9\hsize
\centerline{\epsfbox{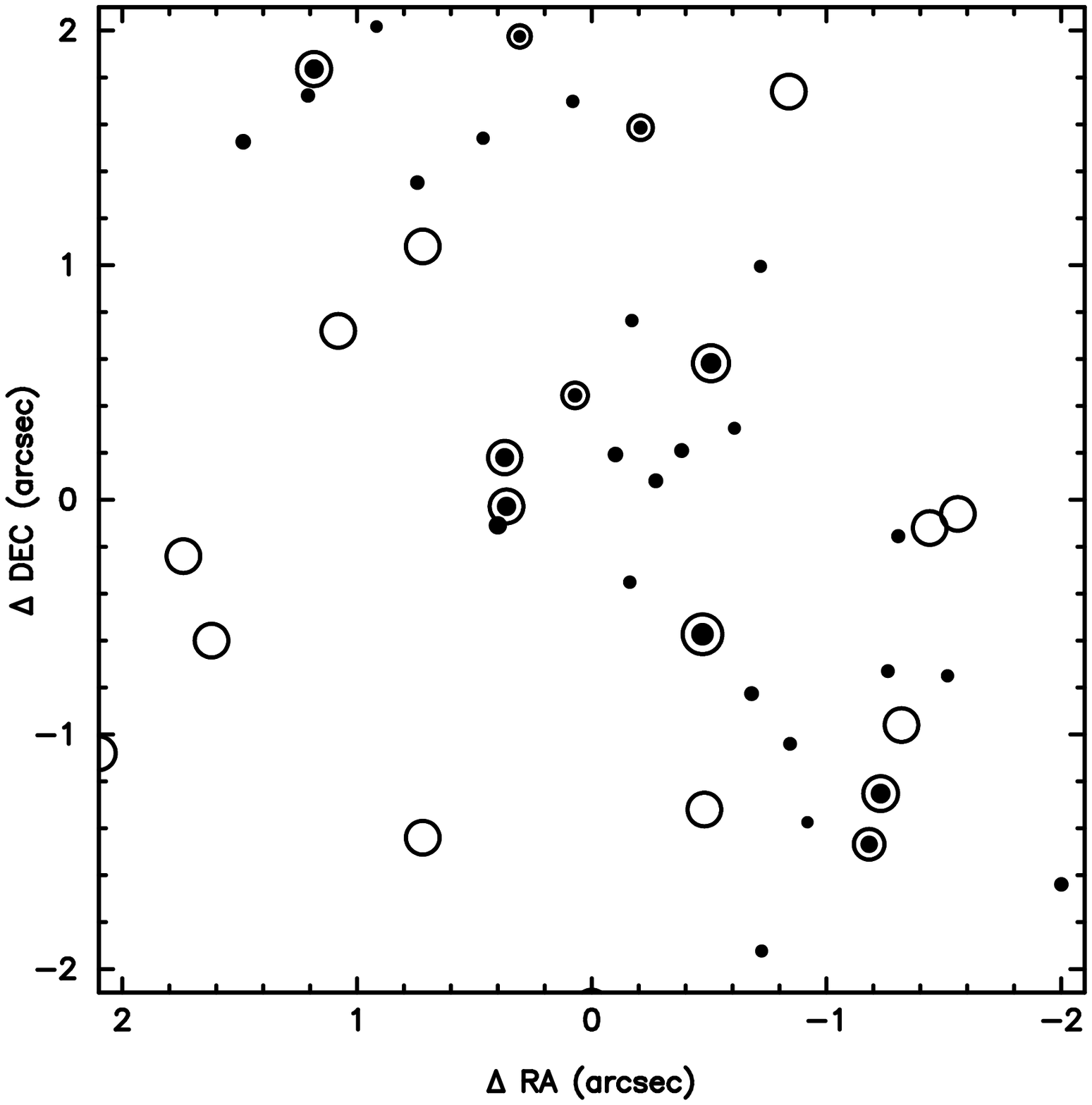}}
\ifsubmode
\vskip3.0truecm
\addtocounter{figure}{1}
\centerline{Figure~\thefigure}
\else\figcaption{\figcapxyplot}\fi
\end{figure}

%%% FIGURE %%%

\clearpage
\begin{figure}
\epsfxsize=0.9\hsize
\centerline{\epsfbox{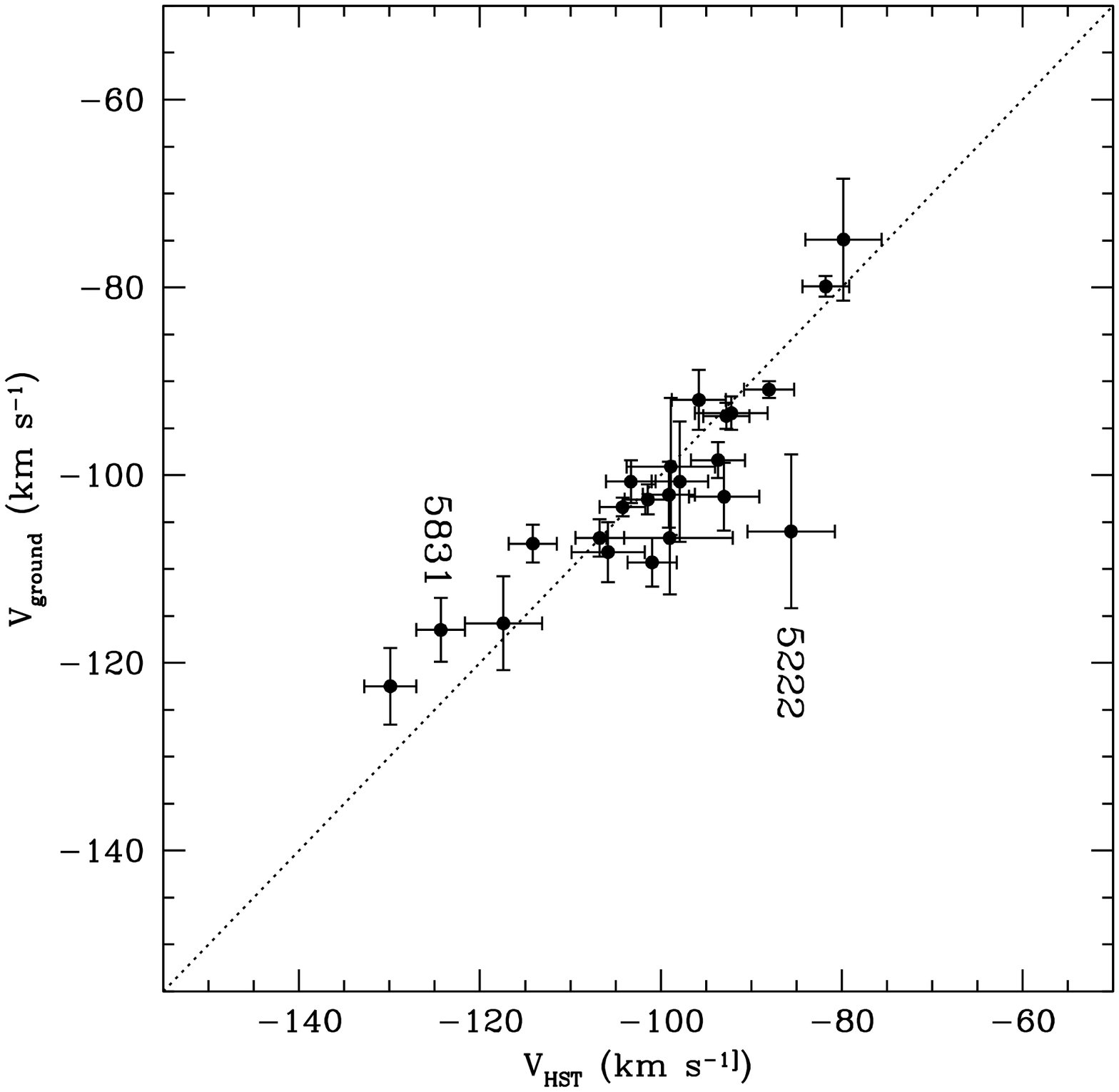}}
\ifsubmode
\vskip3.0truecm
\addtocounter{figure}{1}
\centerline{Figure~\thefigure}
\else\figcaption{\figcapvelcomp}\fi
\end{figure}

%%% FIGURE %%%

\clearpage
\begin{figure}
\epsfxsize=0.9\hsize
\centerline{\epsfbox{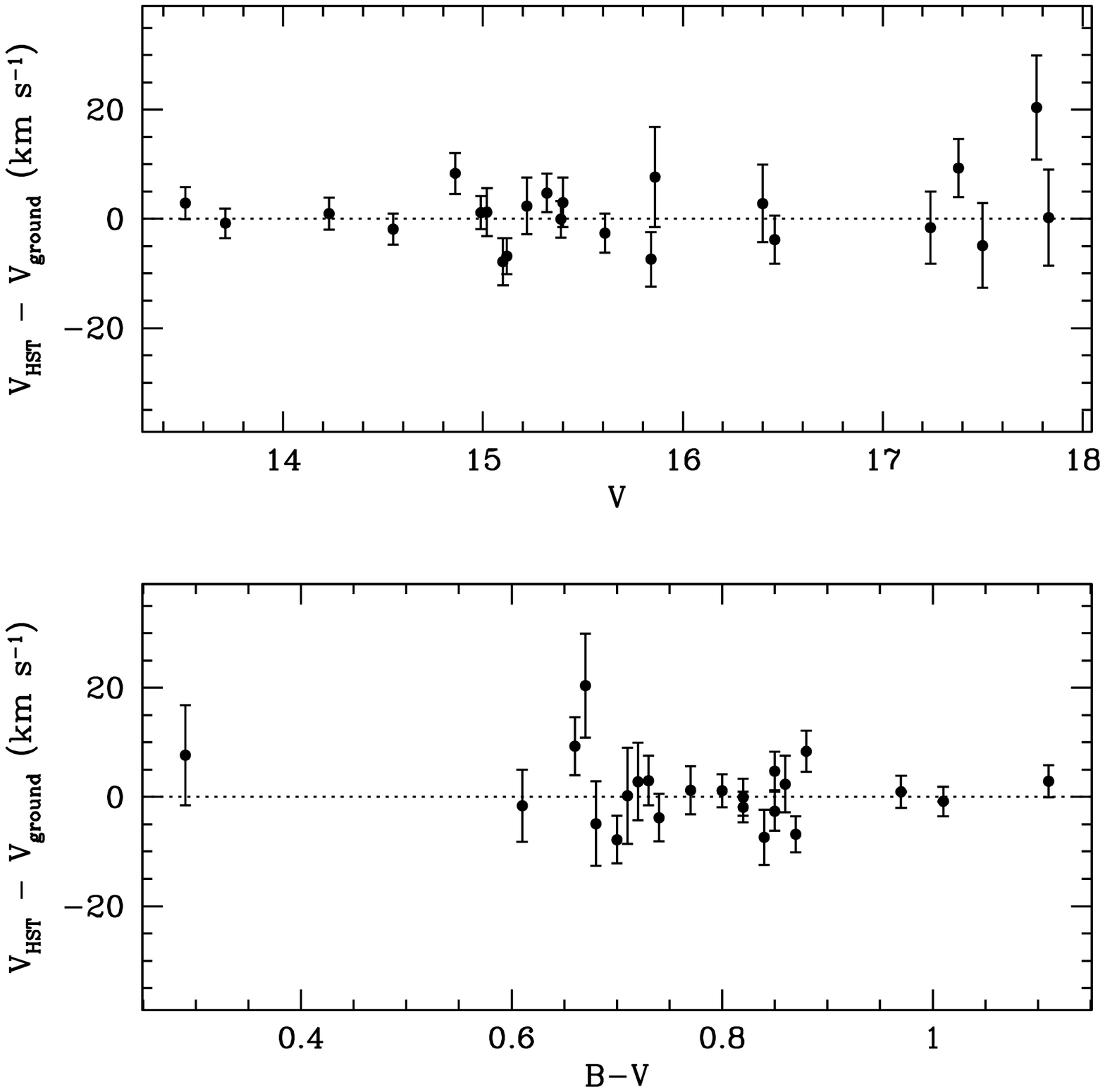}}
\ifsubmode
\vskip3.0truecm
\addtocounter{figure}{1}
\centerline{Figure~\thefigure}
\else\figcaption{\figcapresiduals}\fi
\end{figure}

%%% FIGURE %%%

\clearpage
\begin{figure}
\centerline{\epsfbox{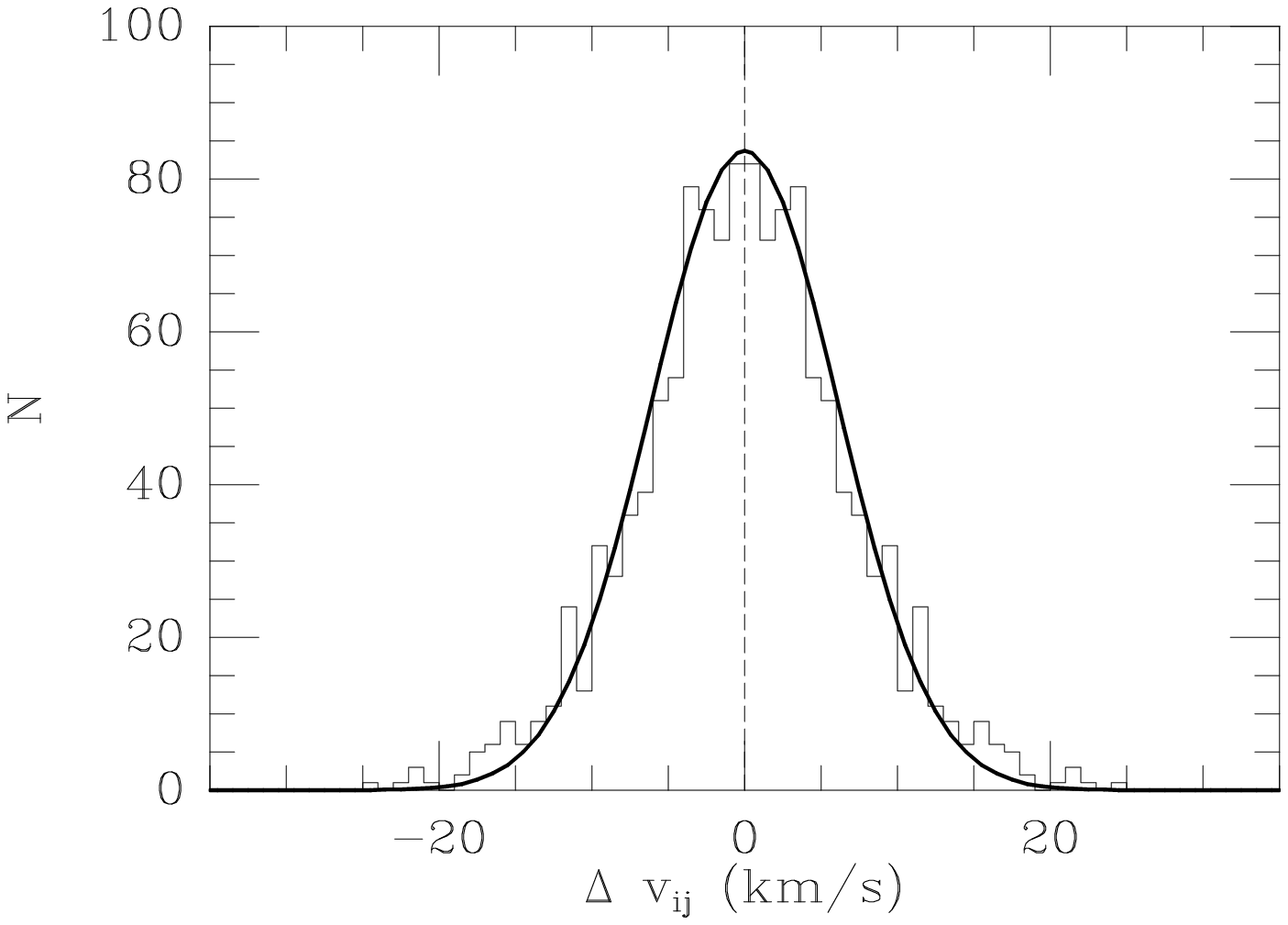}}
\ifsubmode
\vskip3.0truecm
\addtocounter{figure}{1}
\centerline{Figure~\thefigure}
\else\figcaption{\figcaprepeat}\fi
\end{figure}

%%% FIGURE %%%

\clearpage
\begin{figure}
\epsfxsize=0.9\hsize
\centerline{\epsfbox{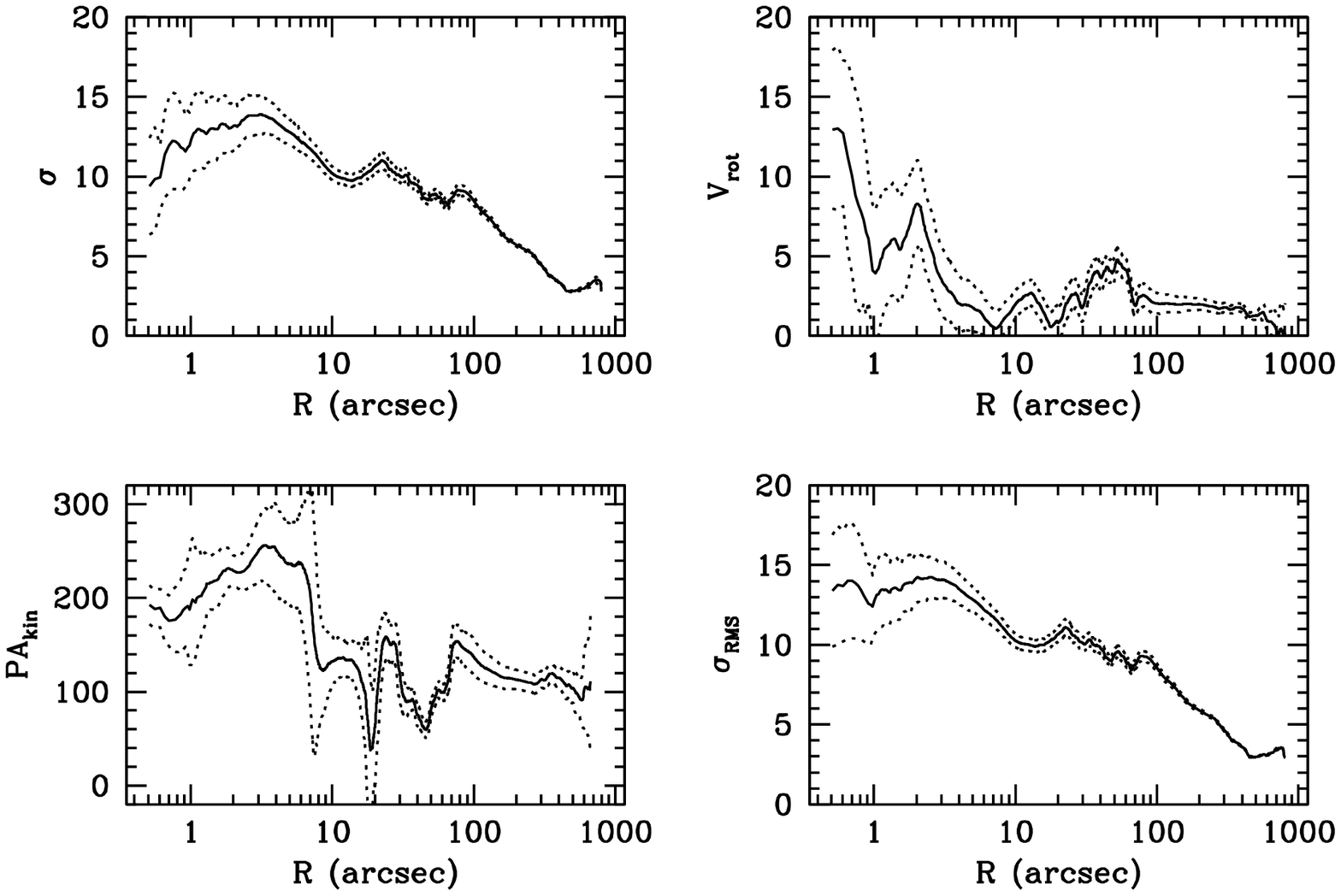}}
\ifsubmode
\vskip3.0truecm
\addtocounter{figure}{1}
\centerline{Figure~\thefigure}
\else\figcaption{\figcapkinprofile}\fi
\end{figure}

%%% FIGURE %%%

\clearpage
\begin{figure}
\epsfxsize=0.6\hsize
\centerline{\epsfbox{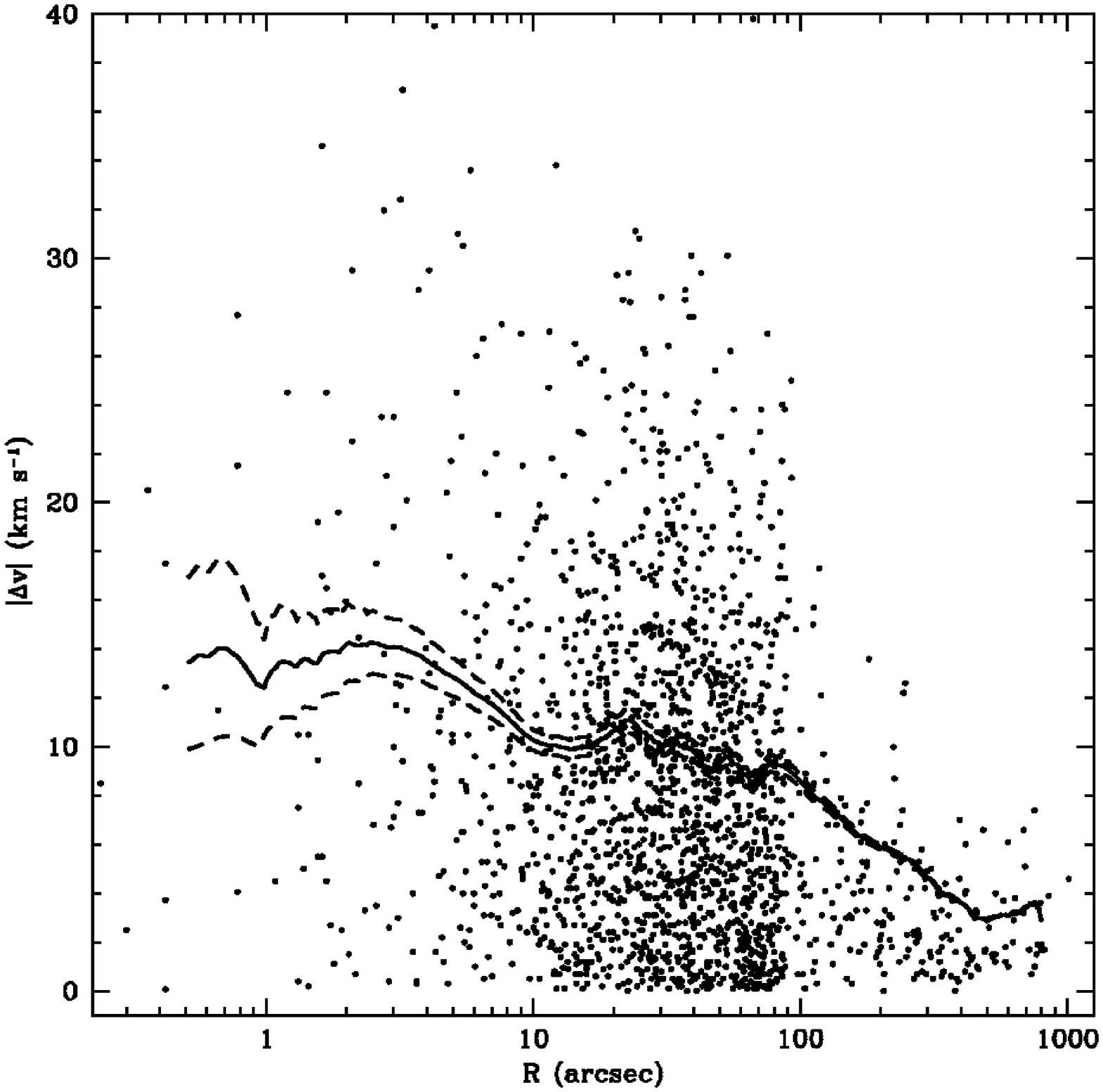}}
\ifsubmode
\vskip3.0truecm
\addtocounter{figure}{1}
\centerline{Figure~\thefigure}
\else\figcaption{\figcaptrend}\fi
\end{figure}

%%% FIGURE %%%

\clearpage
\begin{figure}
\epsfxsize=0.9\hsize
\centerline{\epsfbox{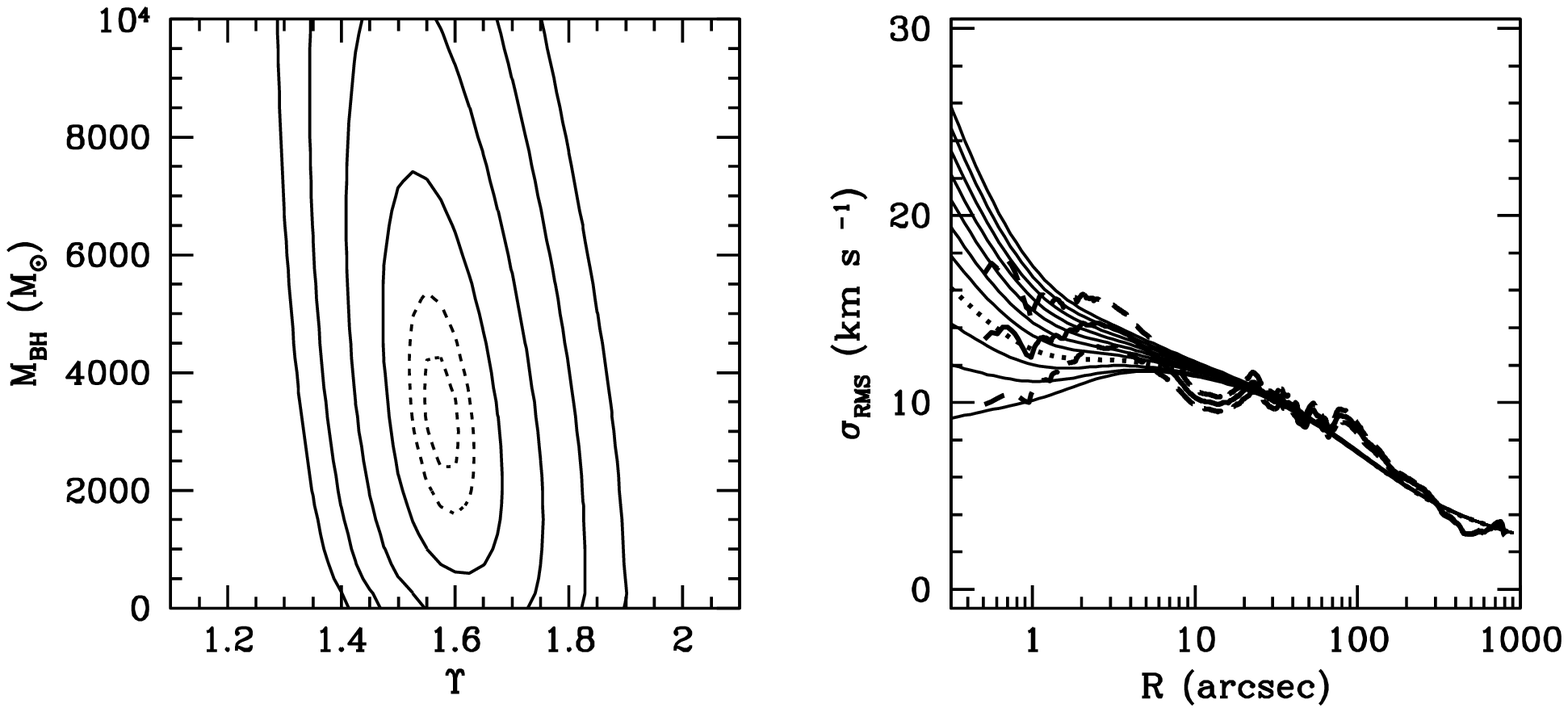}}
\ifsubmode
\vskip3.0truecm
\addtocounter{figure}{1}
\centerline{Figure~\thefigure}
\else\figcaption{\figcapmodels}\fi
\end{figure}

%%% FIGURE %%%

\clearpage
\begin{figure}
\epsfxsize=0.9\hsize
\centerline{\epsfbox{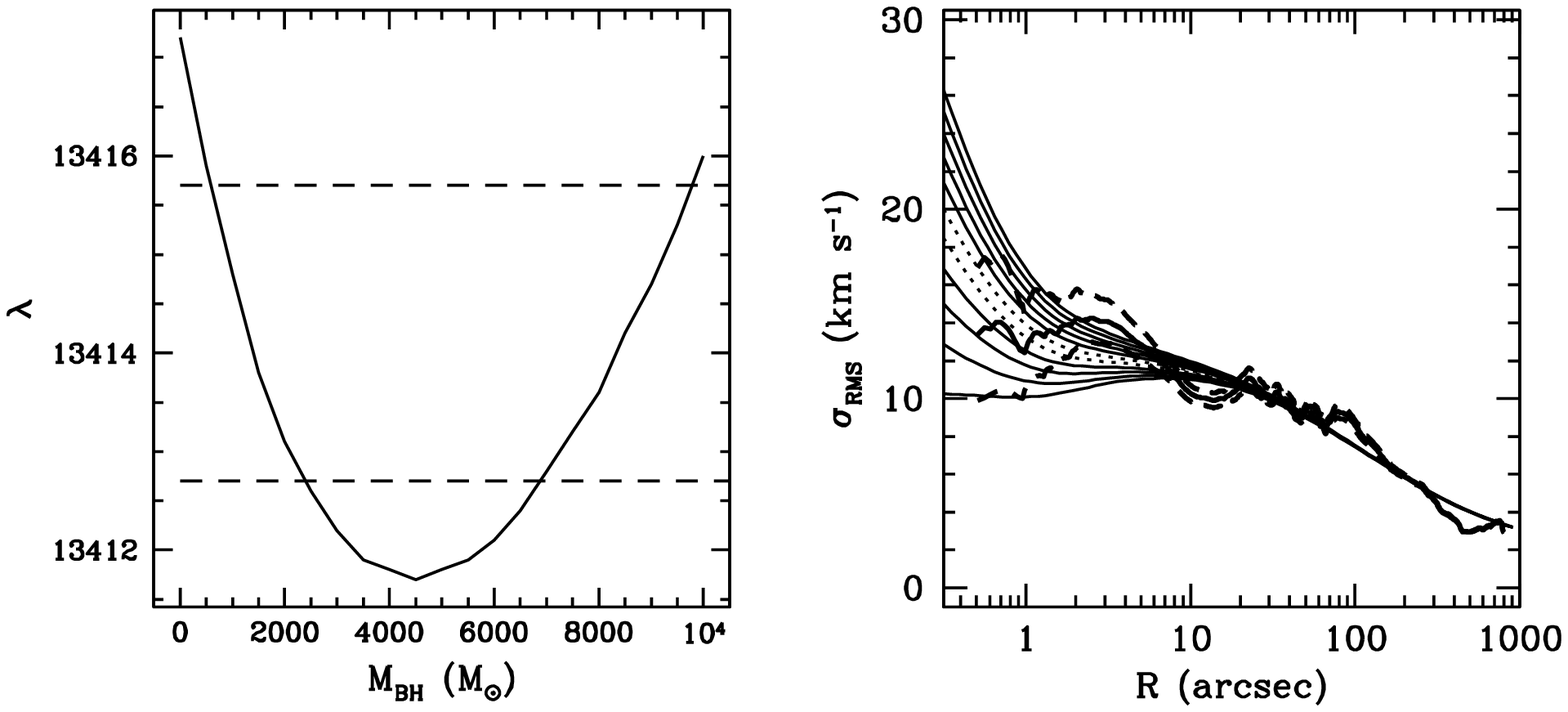}}
\ifsubmode
\vskip3.0truecm
\addtocounter{figure}{1}
\centerline{Figure~\thefigure}
\else\figcaption{\figcapmaxlike}\fi
\end{figure}

%%% FIGURE %%%

\clearpage
\begin{figure}
\epsfxsize=0.9\hsize
\centerline{\epsfbox{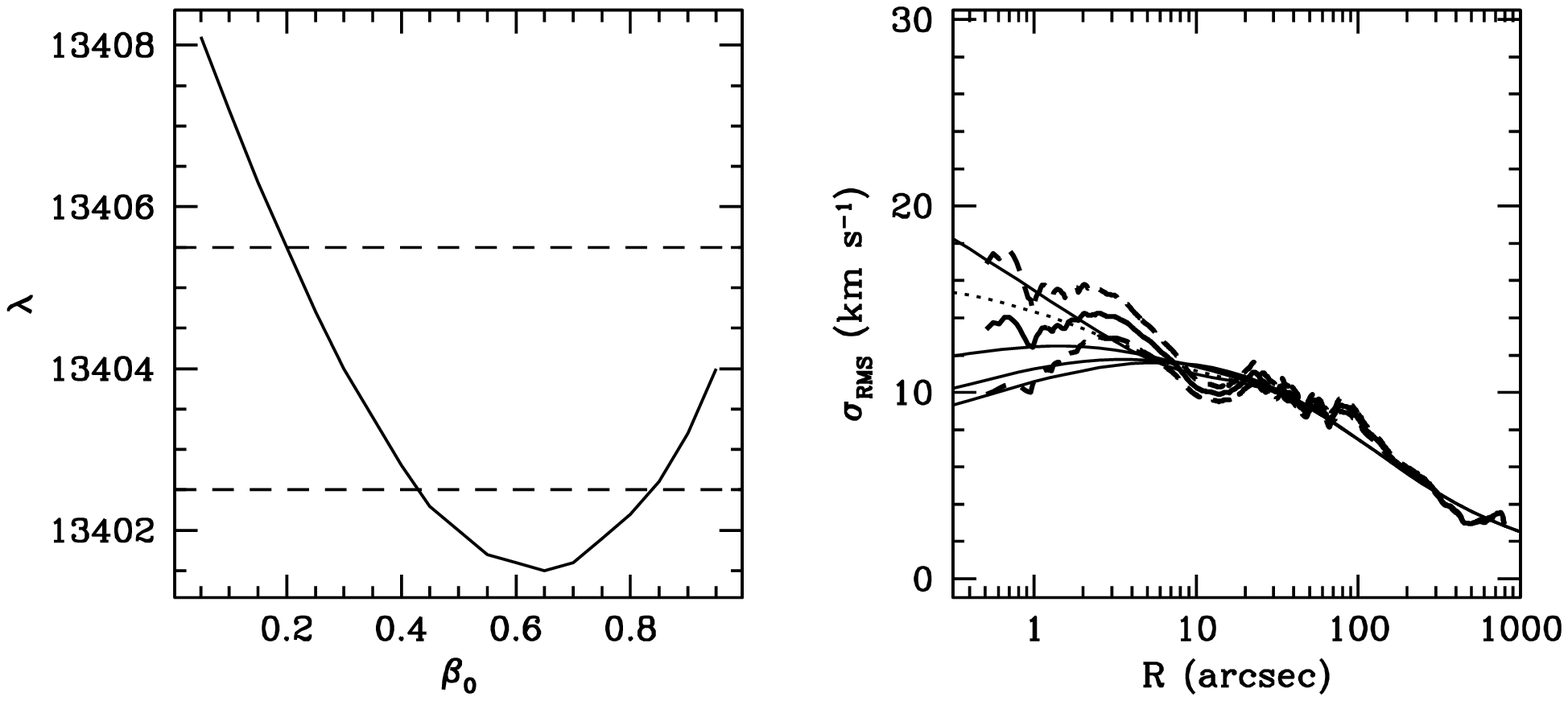}}
\ifsubmode
\vskip3.0truecm
\addtocounter{figure}{1}
\centerline{Figure~\thefigure}
\else\figcaption{\figcapanisotropy}\fi
\end{figure}

%%% FIGURE %%%

\clearpage
\begin{figure}
\epsfxsize=0.9\hsize
\centerline{\epsfbox{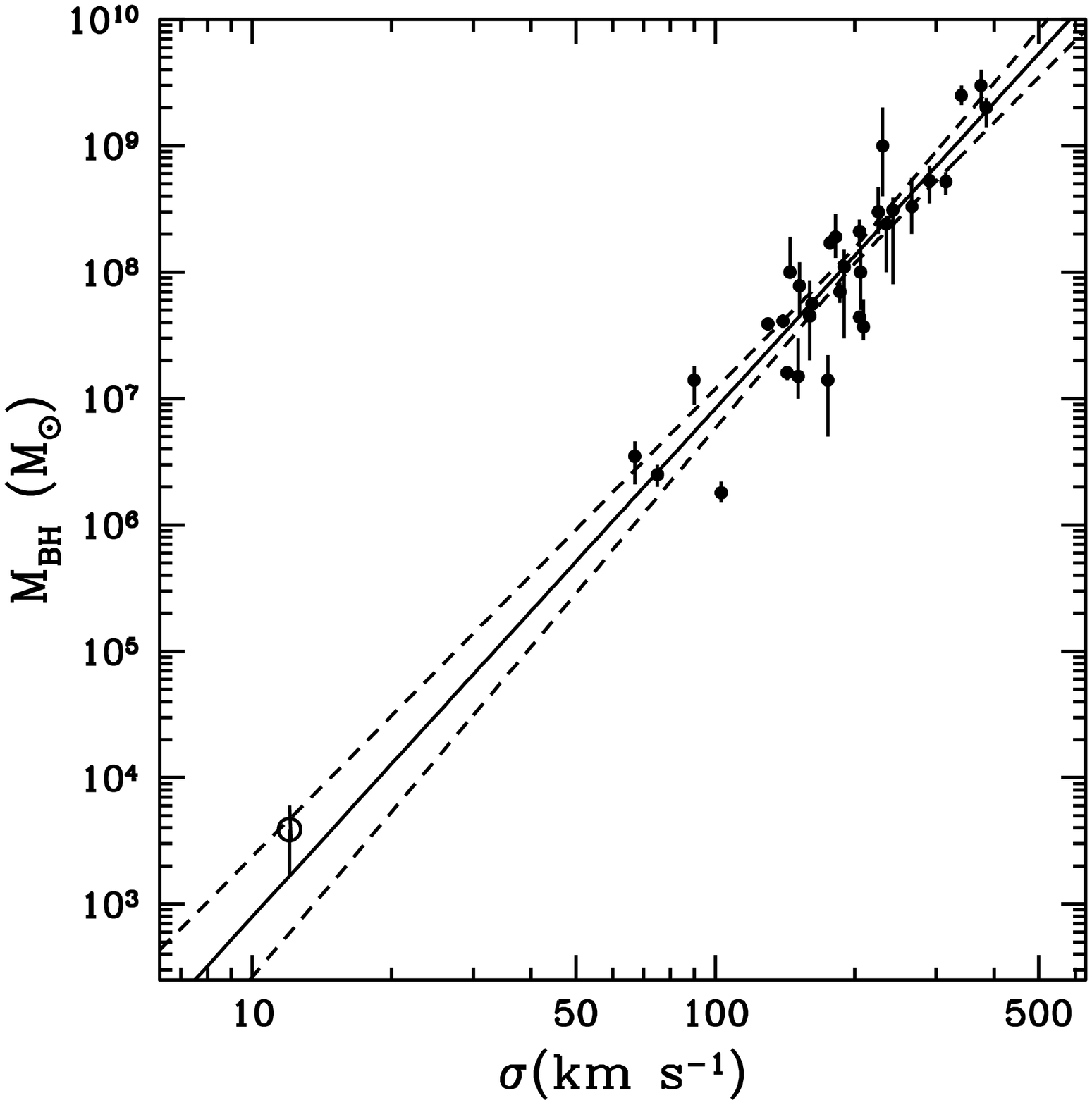}}
\ifsubmode
\vskip3.0truecm
\addtocounter{figure}{1}
\centerline{Figure~\thefigure}
\else\figcaption{\figcapbhcorr}\fi
\end{figure}

%%% END OF FIGURES %%%

\fi % end of \ifprintfig

%%%%%%%%%%%%%%%
% Tables
%%%%%%%%%%%%%%%

\clearpage
\ifsubmode\pagestyle{empty}\fi

\begin{deluxetable}{rrrrrrrrrrrrr}
%\begin{deluxetable}{lcccccccccccc}
\footnotesize
\tiny
\tablewidth{0pc}
\tablecaption{HST/STIS stellar velocity results\label{t:results}}
\tablehead{
\colhead{ID} & \colhead{$\Delta$ RA} & \colhead{$\Delta$ DEC} &
\colhead{$V$} & \colhead{$B-V$} & \colhead{$v_{\rm obs}$} &
\colhead{$\Delta v_{\rm obs}$} & \colhead{$r_{\rm cc}$} & \colhead{$f$} &
\colhead{$v_\star$} & \colhead{$\Delta v_\star$} &
\colhead{$v_{\rm ground}$} & \colhead{$\Delta v_{\rm ground}$} \\
\colhead{} & \colhead{(arcsec)} & \colhead{(arcsec)} & \colhead{} &
\colhead{} & \colhead{(km s$^{-1}$)} & \colhead{(km s$^{-1}$)} &
\colhead{} & \colhead{} & \colhead{(km s$^{-1}$)} & \colhead{(km s$^{-1}$)} &
\colhead{(km s$^{-1}$)} & \colhead{(km s$^{-1}$)} \\
\colhead{(1)} & \colhead{(2)}  & \colhead{(3)} & \colhead{(4)} &
\colhead{(5)} & \colhead{(6)}  & \colhead{(7)} & \colhead{(8)} &
\colhead{(9)} & \colhead{(10)} & \colhead{(11)}  & \colhead{(12)} &
\colhead{(13)}}
\ifsubmode\renewcommand{\arraystretch}{0.68}\fi
\startdata
   696 &   7.44 &  14.15 & 18.62 &   0.53 &--102.7 & 5.0 &   2.06 &  0.836 &--101.8 & 6.5 &        &     \\
   824 &   6.90 &  13.95 & 17.83 &   0.71 & --98.9 & 4.9 &   2.13 &  0.932 & --98.3 & 5.3 & --99.1 & 7.3 \\
  1299 &   6.28 &  11.93 & 19.05 &   0.48 &--116.2 & 4.6 &   2.88 &  0.867 &--117.6 & 5.6 &        &     \\
  2341 &   4.21 &   8.76 & 17.45 &   0.71 &--107.6 & 3.5 &   4.34 &  0.889 &--107.6 & 4.2 &        &     \\
  2357 &   4.34 &   8.51 & 14.99 &   0.80 &--101.4 & 2.6 &  20.82 &  0.991 &--101.4 & 2.6 &--102.6 & 1.6 \\
  2703 &   3.95 &   7.23 & 13.51 &   1.11 & --88.0 & 2.8 &  10.59 &  0.995 & --87.9 & 2.8 & --90.9 & 0.9 \\
  3393 &   2.82 &   5.66 & 15.86 &   0.29 & --99.0 & 7.0 &   2.22 &  0.984 & --98.9 & 7.1 &--106.7 & 6.0 \\
  3726 &   2.38 &   4.91 & 18.03 &   0.69 & --81.1 & 6.3 &   2.35 &  0.893 & --77.9 & 7.2 &        &     \\
  3798 &   2.52 &   4.48 & 17.62 &   0.71 & --88.1 & 3.3 &   4.31 &  0.809 & --83.5 & 5.0 &        &     \\
  3911 &   2.32 &   4.36 & 15.32 &   0.85 & --93.7 & 3.0 &   6.80 &  0.928 & --92.6 & 3.3 & --98.4 & 1.9 \\
  4185 &   1.48 &   4.24 & 18.41 &   0.61 & --98.9 & 3.7 &   2.08 &  0.752 & --96.0 & 6.3 &        &     \\
  4768 &   0.95 &   2.86 & 17.63 &   0.69 & --88.3 & 4.4 &   2.75 &  0.835 & --84.5 & 5.7 &        &     \\
  4891 &   0.77 &   2.65 & 17.50 &   0.68 & --79.8 & 4.2 &   3.38 &  0.897 & --76.6 & 4.9 & --74.9 & 6.5 \\
  4908 &   0.95 &   2.40 & 18.50 &   0.57 &--121.2 & 4.2 &   2.19 &  0.751 &--125.7 & 6.9 &        &     \\
  5002 &   1.18 &   1.84 & 15.02 &   0.77 & --92.2 & 4.0 &   2.73 &  0.903 & --90.5 & 4.6 & --93.4 & 1.8 \\
  5003 &   1.49 &   1.53 & 16.60 &   0.71 &--134.9 & 3.0 &   6.99 &  0.911 &--137.5 & 3.5 &        &     \\
  5029 &   1.21 &   1.72 & 17.10 &   0.70 & --90.9 & 3.7 &   4.59 &  0.754 & --85.5 & 6.2 &        &     \\
  5031 &   0.92 &   2.02 & 17.83 &   0.79 &--114.4 & 3.6 &   3.61 &  0.784 &--116.3 & 5.7 &        &     \\
  5222 &   0.31 &   1.98 & 17.77 &   0.67 & --85.6 & 4.8 &   2.83 &  0.753 & --78.4 & 7.5 &--106.0 & 8.2 \\
  5263 &   0.74 &   1.35 & 17.03 &   0.74 &--112.3 & 3.4 &   4.96 &  0.861 &--113.0 & 4.4 &        &     \\
  5304 &   0.46 &   1.54 & 17.53 &   0.66 &--111.7 & 4.9 &   2.09 &  0.753 &--113.1 & 7.6 &        &     \\
  5380 &   0.08 &   1.70 & 17.48 &   0.68 & --88.6 & 5.3 &   2.48 &  0.778 & --83.2 & 7.6 &        &     \\
  5515 & --0.21 &   1.59 & 17.24 &   0.61 &--117.4 & 4.3 &   3.40 &  0.799 &--119.9 & 6.1 &--115.8 & 5.0 \\
  5755 & --0.17 &   0.76 & 17.48 &   0.62 &--124.0 & 7.9 &   2.20 &  0.765 &--129.1 &11.0 &        &     \\
  5768 &   0.37 &   0.18 & 15.22 &   0.86 &--105.8 & 4.1 &   3.52 &  0.854 &--105.5 & 5.2 &--108.2 & 3.2 \\
  5785 &   0.07 &   0.44 & 17.02 &   0.76 &--109.9 & 3.1 &   6.79 &  0.753 &--110.7 & 5.7 &        &     \\
  5831 &   0.36 & --0.03 & 15.10 &   0.70 &--124.3 & 2.7 &  12.53 &  0.797 &--128.6 & 4.6 &--116.5 & 3.4 \\
  5846 &   0.40 & --0.11 & 15.49 &   0.34 & --70.2 & 3.7 &   4.24 &  0.755 & --58.1 & 6.3 &        &     \\
  5864 & --0.72 &   1.00 & 17.61 &   0.53 &--126.9 & 4.0 &   2.55 &  0.765 &--132.9 & 6.4 &        &     \\
  5926 & --0.10 &   0.19 & 16.61 &   0.26 &--101.3 & 3.4 &   4.15 &  0.799 & --99.8 & 5.3 &        &     \\
  5933 & --0.51 &   0.58 & 14.55 &   0.82 & --81.8 & 2.6 &  17.68 &  0.934 & --79.9 & 2.9 & --79.9 & 1.1 \\
  6005 & --0.38 &   0.21 & 16.89 &   0.57 & --94.2 & 3.7 &   3.88 &  0.791 & --90.6 & 5.7 &        &     \\
  6012 & --0.27 &   0.08 & 16.88 &   0.68 &--109.7 & 3.3 &   5.85 &  0.761 &--110.4 & 5.7 &        &     \\
  6044 & --0.61 &   0.31 & 17.66 &   0.60 & --98.9 & 4.9 &   2.49 &  0.759 & --96.2 & 7.6 &        &     \\
  6111 & --0.16 & --0.35 & 17.49 &   0.72 & --92.0 & 3.5 &   4.04 &  0.761 & --87.1 & 5.9 &        &     \\
  6290 & --0.47 & --0.57 & 13.71 &   1.01 &--104.2 & 2.5 &  35.00 &  0.984 &--104.2 & 2.6 &--103.4 & 1.0 \\
  6433 & --1.31 & --0.16 & 17.19 &   0.45 & --99.0 & 3.7 &   3.37 &  0.820 & --97.1 & 5.2 &        &     \\
  6450 & --0.68 & --0.83 & 16.90 &   0.01 &--111.2 & 4.7 &   2.11 &  0.752 &--112.4 & 7.4 &        &     \\
  6575 & --0.85 & --1.04 & 17.33 &   0.62 &--113.7 & 3.5 &   4.82 &  0.786 &--115.3 & 5.5 &        &     \\
  6617 & --1.26 & --0.73 & 17.33 &   0.73 &--117.0 & 4.3 &   4.04 &  0.846 &--118.7 & 5.5 &        &     \\
  6711 & --1.52 & --0.75 & 17.57 &   0.74 & --95.1 & 6.5 &   2.02 &  0.779 & --91.6 & 9.0 &        &     \\
  6719 & --0.92 & --1.37 & 17.95 &   0.75 &--104.2 & 3.8 &   2.87 &  0.780 &--103.2 & 6.0 &        &     \\
  6772 & --1.23 & --1.25 & 14.86 &   0.88 &--101.0 & 2.7 &  11.57 &  0.949 &--100.6 & 2.9 &--109.3 & 2.6 \\
  6828 & --0.72 & --1.92 & 17.64 &   0.63 &--109.4 & 5.2 &   2.00 &  0.830 &--109.7 & 6.8 &        &     \\
  6833 & --1.18 & --1.47 & 15.84 &   0.84 &--129.9 & 2.9 &   9.86 &  0.856 &--133.7 & 3.9 &--122.5 & 4.1 \\
  7148 & --2.00 & --1.64 & 17.05 &   0.70 &--111.4 & 3.4 &   4.43 &  0.913 &--111.8 & 3.9 &        &     \\
  7313 & --2.02 & --2.19 & 15.40 &   0.73 & --99.1 & 2.9 &   8.79 &  0.928 & --98.4 & 3.3 &--102.1 & 3.5 \\
  7404 & --1.74 & --2.78 & 16.40 &   0.72 & --97.9 & 3.1 &   6.34 &  0.958 & --97.5 & 3.3 &--100.7 & 6.4 \\
  7656 & --1.36 & --4.00 & 17.38 &   0.66 & --93.0 & 3.9 &   2.64 &  0.831 & --90.0 & 5.3 &--102.3 & 3.6 \\
  7718 & --2.39 & --3.14 & 14.23 &   0.97 & --92.8 & 2.6 &  23.41 &  0.991 & --92.6 & 2.6 & --93.7 & 1.4 \\
  7775 & --2.05 & --3.76 & 17.65 &   0.66 &--142.2 & 4.8 &   3.41 &  0.876 &--147.2 & 5.8 &        &     \\
  8146 & --1.89 & --5.19 & 17.30 &   0.16 &--120.5 & 4.1 &   2.08 &  0.833 &--123.1 & 5.5 &        &     \\
  8194 & --3.09 & --4.13 & 15.61 &   0.85 &--103.3 & 2.7 &  11.20 &  0.975 &--103.2 & 2.8 &--100.7 & 2.3 \\
  8260 & --3.10 & --4.41 & 17.33 &   0.69 &--102.6 & 3.9 &   3.89 &  0.851 &--101.7 & 5.0 &        &     \\
  8292 & --3.04 & --4.57 & 17.57 &   0.66 &--102.7 & 8.5 &   2.27 &  0.824 &--101.7 &10.6 &        &     \\
  8362 & --2.91 & --4.93 & 18.22 &   0.62 &--115.9 & 4.3 &   2.09 &  0.787 &--118.2 & 6.3 &        &     \\
  8396 & --3.13 & --4.83 & 17.28 &   0.67 &--105.5 & 3.4 &   4.07 &  0.907 &--105.3 & 3.9 &        &     \\
  8492 & --3.31 & --4.97 & 17.60 &   0.00 &--111.5 & 4.9 &   2.12 &  0.909 &--111.9 & 5.6 &        &     \\
  8496 & --2.86 & --5.45 & 16.46 &   0.74 & --95.8 & 3.0 &   6.72 &  0.941 & --95.1 & 3.3 & --92.0 & 3.2 \\
  8760 & --2.86 & --6.38 & 17.72 &   0.75 & --96.6 & 3.8 &   4.05 &  0.863 & --94.9 & 4.8 &        &     \\
  8917 & --3.14 & --6.68 & 16.84 &   0.70 & --89.8 & 3.9 &   2.41 &  0.933 & --88.5 & 4.3 &        &     \\
 10602 & --6.73 &--11.08 & 15.39 &   0.82 &--106.8 & 2.7 &  13.06 &  0.996 &--106.8 & 2.7 &--106.7 & 2.0 \\
 14442 &  11.48 &  22.89 & 17.48 &   0.68 &--109.5 & 4.7 &   3.19 &  0.990 &--109.5 & 4.7 &        &     \\
 14526 &  12.07 &  24.37 & 15.12 &   0.87 &--114.1 & 2.7 &  13.37 &  0.999 &--114.2 & 2.7 &--107.3 & 2.0 \\
\enddata
\tablecomments{The ID number in column (1) corresponds to the number in the
M15 stellar catalog presented in Paper~I. The stellar positions
[column (2) and (3)], $V$-band magnitudes [column (4)] and $B-V$
colors [column (5)] are taken from this catalog. The line-of-sight
velocities inferred from the STIS spectra, and their uncertainties,
are listed in columns (6) and (7), respectively. The $r_{\rm cc}$
cross-correlation statistic associated with the inferred velocity (see
Section~\ref{s:extraction}) is listed in column (8). The quantity $f$
in column (9) measures the amount of blending in the spectrum (see
Section~\ref{ss:blending}). The blending corrected velocities and
their errors are shown in columns (10) and (11). Ground-based velocity
determinations and their associated errors, from the compilation of
Gebhardt \etal (2000a), are listed in columns (12) and
(13). Ground-based velocity determinations with uncertainties in
excess of $10 \kms$ were omitted (these were available for stars 5785
and 8760).}
\end{deluxetable}

%%%%%%%%%%%%%%%
% End of Document
%%%%%%%%%%%%%%%


\clearpage

\begin{thebibliography}{}

\bibitem{And00}
Anderson, J., \& King, I. R. 2000, PASP, 112, 1360

\bibitem{Bah76}
Bahcall, J. N., \& Wolf R. A. 1976, ApJ, 209, 214

\bibitem{Bah77}
Bahcall, J. N., \& Wolf R. A. 1977, ApJ, 216, 883

\bibitem{Bau02}
Baumgardt, H., Heggie, D. C., \& Hut, P. 2002, MNRAS, in press 
[astro-ph/0206258] 

\bibitem{Bin82}
Binney, J. J., \& Mamon G. A. 1982, MNRAS, 200, 361

\bibitem[]{Bin98}
Binney, J. J., \& Merrifield, M. 1998, Galactic Astronomy
(Princeton: Princeton University Press)

\bibitem[]{Bin87}
Binney, J., \& Tremaine, S. 1987, Galactic Dynamics
(Princeton: Princeton University Press)

\bibitem[]{Col99}
Colbert, E. J. M., \& Mushotzky, R. F. 1999, ApJ, 519, 89

\bibitem[]{dAm02}
d'Amico, N., Possenti, A., Fici, L., Manchester, R. N., Lyne, A. G.,
Camilo, F., \& Sarkissian, J. 2002, ApJL, 570, L89

\bibitem[]{Dav98}
Davies, M., \& Hansen, B. 1998, MNRAS, 301, 15

\bibitem[]{deM95}
De Marchi, G., \& Paresce, F. 1995, A\&A, 304, 202

\bibitem{Djo86}
Djorgovski, S., \& King, I. 1986, ApJ, 305, 61

\bibitem{Dru96}
Drukier, G. A. 1996, MNRAS, 280, 498

\bibitem{Dru98}
Drukier, G. A., Slavin, S. D., Cohn, H. N., Lugger, P. M., Berrington, R.C.,
Murphy, B. W., \& Seitzer, P. O. 1998, AJ, 115, 708

\bibitem{Dub94b}
Dubath, P., Meylan, G., \& Mayor, M. 1994, ApJ, 426, 192

\bibitem{Dul97}
Dull, J. D., Cohn, H. N., Lugger, P. M., Murphy, B. W., Seitzer, P. O.,
Callanan, P. J., Rutten, R. G. M., \& Charles, P. A. 1997, ApJ, 481, 267

\bibitem[]{Fer00}
Ferrarese, L., \& Merritt, D. 2000, ApJ, 539, L9

\bibitem[]{Geb94}
Gebhardt, K., Pryor, C., Williams, T. B., \&
Hesser, J. E. 1994, AJ, 107, 2067

\bibitem[]{Geb95}
Gebhardt, K., \& Fischer, P. 1995, AJ, 109, 209

\bibitem[]{Geb00a}
Gebhardt, K., Pryor, C., O'Connell, R. D., Williams, T. B., \&
Hesser, J. E. 2000a, AJ, 119, 1268

\bibitem[]{Geb00b}
Gebhardt, K., et al. 2000b, ApJ, 539, L13

\bibitem[]{Geb02}
Gebhardt, K., Rich, R. M., \& Ho, L. 2002, ApJL, in press

\bibitem[]{Ger93}
Gerhard, O. E. 1993, MNRAS, 265, 213

\bibitem{Gra92}
Grabhorn, R. P., Cohn, H. N., Lugger, P. M., \& Murphy, B. W.
1992, ApJ, 392, 86

\bibitem[]{Guh96}
Guhathakurta, P., Yanny, B., Schneider, D. P., \& Bahcall, J. N.,
1996, AJ, 111, 267

\bibitem[]{Han97}
Hansen, B., \& Phinney, E. S. 1997, MNRAS, 291, 569

\bibitem[]{Har96}
Harris, W.E. 1996, AJ, 112, 1487

\bibitem[]{Hut92}
Hut, P., et al. 1992, PASP, 104, 981

\bibitem[]{Kor01}
Kormendy, J., \& Gebhardt, K. 2001, in `Proc. 20th Texas Symposium on
relativistic astrophysics', AIP conference proceedings, Vol. 586.,
eds., J. C. Wheeler, \& H. Martel, p. 363 (NY: American Institute of
Physics)

\bibitem[]{Kro00}
Kronawitter, A., Saglia, R. P., Gerhard, O., Bender, R. 2000, A\&AS, 144, 53

\bibitem{Kur}
Kurtz, M. J., \& Mink, D. J. 1998, PASP, 110, 934

\bibitem[]{Lau91}
Lauer, T. R., et al. 1991, ApJ, 369, L45

\bibitem[]{Lee87}
Lee, H. M. 1987, ApJ, 319, 801

\bibitem[]{Lee93}
Lee, M.~H. 1993, ApJ, 418, 147

\bibitem[]{Lee95}
Lee, H.~M. 1995, MNRAS, 272, 605

\bibitem[]{Lin99}
Lineweaver, C. 1999, Science, 284, 1503

\bibitem[]{lug87}
Lugger, P. M., Cohn, H., Grindlay, J. E., Bailyn, C. D., Hertz, P. 
1987, ApJ, 320, 482

\bibitem[]{May80}
Mayor, M. 1980, A\&A, 87, L1

\bibitem[]{Mer93}
Merritt, D. R., \& Saha, P. 1993, ApJ, 409, 75

\bibitem[]{Mil02}
Miller, M. C., \& Hamilton, D. P. 2002, MNRAS, 330, 232

\bibitem[]{Mou02}
Mouri, H., \& Taniguchi, Y. 2002, ApJ, 566, L17

\bibitem{Mur90}
Murphy, B. W., Cohn, H. N., \& Hut, P. 1990, MNRAS, 245, 335

\bibitem{Mur97}
Murphy, B. W., Cohn, H. N., Lugger, P. M., \& Drukier, G. A. 1997,
BAAS, 29, 1338

\bibitem{Par00}
Paresce, F., \& De Marchi, G. 2000, ApJ, 534, 870

\bibitem[]{Pet93}
Peterson, R. C. 1993, in Structure and Dynamics of Globular
Clusters, eds., G. Djorgovski \& G. Meylan, ASP Conference Series, Vol.~50,
p.~65

\bibitem{Pet89}
Peterson, R. C., Seitzer, P., \& Cudworth, K. M. 1989, ApJ, 347, 251

\bibitem[]{Pfa02}
Pfahl, E., Rappaport, S., \& Podsiadlowski, P. 2002, ApJ, 573, 283

\bibitem[]{Phi93}
Phinney, E. S. 1993, in Structure and Dynamics of Globular Clusters,
eds., G. Djorgovski \& G. Meylan, ASP Conference Series, Vol.~50, p.~141

\bibitem[]{Por02}
Portegies Zwart, S. F., \& McMillan, S. L. W. 2002, ApJ, in press
[astro-ph/0201055]

\bibitem[]{Pre92}
Press, W. H., Teukolsky, S. A., Vetterling, W. T., \& Flannery, B. P. 1992,
Numerical Recipes (Cambridge: Cambridge University Press)

\bibitem[]{Pry89}
Pryor, C., McClure, R. D., Fletcher, J. M., \& Hesser, J. E. 1989,
in Dynamics of Dense Stellar Systems, ed. D. Merritt, p. 175
(Cambridge: Cambridge University Press)

\bibitem{Qui87}
Quinlan, G. D., \& Shapiro, S. L. 1987, ApJ, 321, 199

\bibitem{Qui90}
Quinlan, G. D., \& Shapiro, S. L. 1990, ApJ, 356, 483

\bibitem[]{Ree84}
Rees, M. J. 1984, ARA\&A, 22, 471

\bibitem[]{San70}
Sanders, R. H. 1970, ApJ, 162, 791

\bibitem[]{Smi95}
Smith, H. A. 1995, RR Lyrae Stars (Cambridge: Cambridge University Press)

\bibitem[]{Sne83}
Sneden, C., \& Parthasarathy, M. 1983, ApJ, 267, 757

\bibitem[]{Sos97}
Sosin, C., \& King, I. R. 1997, AJ, 113, 1328

\bibitem[]{Stu91}
Stuart, A., \& Ord, J. K. 1991, Kendall's Advanced Theory of
Statistics, Volume II, 5th ed. (London: Edward Arnold, a division of
Hodder \& Stoughton)

\bibitem[]{Tak96}
Takahashi, K. 1996, PASJ, 48, 691

\bibitem{Ton79}
Tonry, J., \& Davis, M. 1979, AJ, 84, 1511

\bibitem{Tra95}
Trager, S. C., King, I. R., \& Djorgovski, S. 1995, AJ, 109, 218

\bibitem{Tre94}
Tremaine, S., Richstone, D. O., Byun, Y.-I., Dressler, A., Faber, S. M.,
Grillmair, C., Kormendy, J., Lauer, T. R. 1994, AJ, 107, 634

\bibitem{Tre02}
Tremaine, S., et al. 2002, ApJ, 574, 740

\bibitem[]{vdM94}
van der Marel, R. P. 1994, MNRAS, 270, 271

\bibitem[]{vdM01}
van der Marel, R. P. 2001, in `Black Holes in Binaries and Galactic
Nuclei', Kaper L., van den Heuvel E. P. J., Woudt P. A., eds.,
Springer-Verlag, p.~246

\bibitem[]{vdM93}
van der Marel, R. P., \& Franx, M. 1993, ApJ, 407, 525

\bibitem[]{vdM02}
van der Marel, R. P., Gerssen, J., Guhathakurta, P.,
Peterson, R. C., \& Gebhardt, K. 2002, AJ, in press (Paper I)

\bibitem[]{vdM00}
van der Marel R.P., Magorrian J., Carlberg R.G., Yee H.K.C., Ellingson E.
2000, AJ, 119, 2038

\bibitem{Zag93}
Zaggia, S., Capaccioli, M., \& Piotto, G. 1993, A\&A, 278, 415

\bibitem[]{Zez02}
Zezas, A., \& Fabbiano, G. 2002, ApJ, submitted [astro-ph/0203176]

\end{thebibliography}
\end{document}